\documentclass[onecolumn]{aa} 
\usepackage{txfonts}
\usepackage{color}

\usepackage{graphicx}
\usepackage{txfonts}
\begin{document}

   \title{Merging binary black holes formed through double-core evolution}
      \author{Y. Qin\inst{1,2}
          \and
          R.-C. Hu\inst{2}
         \and 
         G. Meynet\inst{3,4}
          \and
         Y. Z. Wang\inst{5}
          \and
         J.-P. Zhu\inst{6}
           \and
         H. F. Song\inst{7}
          \and
         X. W. Shu\inst{1}
          \and
         S. C. Wu\inst{8,9}
          }

   \institute{
            Department of Physics, Anhui Normal University, Wuhu, Anhui, 241000, China\\
              \email{yingqin2013@hotmail.com}
         \and
            Guangxi Key Laboratory for Relativistic Astrophysics, School of Physical Science and Technology, Guangxi University, Nanning 530004, China
        \and  
           Département d’Astronomie, Université de Genève, Chemin Pegasi 51, CH-1290 Versoix, Switzerland
        \and
          Gravitational Wave Science Center (GWSC), Université de Genève, CH-1211 Geneva, Switzerland\\
            \email{Georges.Meynet@unige.ch}
        \and 
            Key Laboratory of Dark Matter and Space Astronomy, Purple Mountain Observatory, Chinese Academy of Sciences, Nanjing, 210033, People's Republic of China
        \and 
            Department of Astronomy, School of Physics, Peking University, Beijing 100871, China            
        \and      
            College of Physics, Guizhou University, Guiyang city, Guizhou Province, 550025, P.R. China 
         \and 
            Max-Planck-Institut f{\"u}r Gravitationsphysik (Albert-Einstein-Institut), D-30167 Hannover, Germany
         \and 
            Leibniz Universit{\"a}t Hannover, D-30167 Hannover, Germany
             }



 \abstract
 {To date, various formation channels of merging events have been heavily explored with the detection of nearly 100 double black hole (BH) merger events reported by the LIGO-Virgo-KAGRA (LVK) Collaboration. We here systematically investigate an alternative formation scenario, i.e., binary BHs (BBHs) formed through double helium stars (hereafter double-core evolution channel). In this scenario, the two helium stars (He-rich stars) could be the outcome of the classical isolated binary evolution scenario involving with and without common-envelope phase (i.e., CE channel and stable mass transfer channel), or alternatively of massive close binaries evolving chemically homogeneously (i.e., CHE channel).}
   {We study the properties (i.e., the chirp masses and the effective spins) of binary BHs (BBHs) formed through the double-core evolution, and investigate the impact of different efficiencies of angular momentum transport within massive He-rich stars on double-core evolution.}
   {We perform detailed stellar structure and binary evolution calculations that take into account internal differential rotation and mass loss of He-rich stars, as well as tidal interactions in binaries. We systematically study the parameter space of initial binary He-rich stars, including initial mass and metallicity of He-rich stars, as well as initial orbital periods. Apart from direct core collapse with mass and angular momentum conserved, we also follow the framework in \citet{2019arXiv190404835B} to estimate the mass and spin of the resulting BHs.}
   {We show that the radii of massive He-rich stars decrease as a function of time, which comes mainly from mass loss and mixing in high metallicity and from mixing in low metallicity. For double He-rich stars with equal masses in binaries, we find that tides start to be at work on the Zero Age Helium Main Sequence (ZAHeMS: the time when a He-rich star starts to burn helium in the core, which is analogous to ZAMS for core hydrogen burning) for initial orbital periods not longer than 1.0 day, depending on the initial metallicities. Besides the stellar mass loss rate and tidal interactions in binaries, we find that the role of the angular momentum transport efficiency in determining the resulting BH spins, becomes stronger when considering BH progenitors originated from a higher metal-metallicity environment. We highlight that double-core evolution scenario does not always produce fast-spinning BBHs and compare the properties of the BBHs reported from the LVK with our modeling.
    }
   {After detailed binary calculations of double-core evolution, we have confirmed that the spin of the BH is not only determined by the interplay of the binary's different initial conditions (metallicity, mass and orbital period), but also dependent on the angular momentum transport efficiency within its progenitor. We predict that, with the sensitivity improvements to the LVK's next observing run (O4), the sample of merging BBHs will contain more sources with positive but moderate (even high) $\chi_{\rm eff}$ and part of the events are likely formed through the double-core evolution channel.}

  \maketitle
%
\section{Introduction}
The LIGO-Virgo-KAGRA (LVK) Collaboration has released the Gravitational Wave Transient Catalog 3 \citep[GWTC-3,][]{2021arXiv211103606T}, consisting of 69 confident binary black hole (BBH) merger events with the detection threshold to count events with false alarm rate (FAR) < 1 $yr^{-1}$. With the targeted sample of BBHs in GWTC-3, the LVK Collaboration has also inferred their intrinsic properties (e.g., merger rates, masses and effective inspiral spins), among which the effective inspiral spin $\chi_{\rm eff}$ \footnote{$\chi_{\rm eff}= (M_1 \chi_{1z} + M_2 \chi_{2z})/(M_1 + M_2)$, where $M_1$ and $M_2$ are the component masses of the two BHs, $\chi_{1z}$ and $\chi_{2z}$ are dimensionless BH spin magnitudes aligned to the direction of the orbital angular momentum (AM).} has been widely considered as a probe to distinguish the formation channels of merging BBH events \citep{2016ApJ...818L..22A,2017Natur.548..426F,2018ApJ...854L...9F,2021arXiv211103634T,2021PhRvD.104h3010R}. The majority of the BBHs reported by the LVK Collaboration have low $\chi_{\rm eff}$, while several BBH mergers \footnote{GW190403 and GW190805 with high $\chi_{\rm eff}$ were reported from deeper searches in GWTC-2.1 \citep{2021arXiv210801045T}, but a low-significance FAR threshold of 2 per day.} show definitely high positive $\chi_{\rm eff}$, e.g., $0.28_{-0.29}^{+0.26}$, $0.31_{-0.22}^{+0.20}$, $0.33_{-0.25}^{+0.22}$, $0.37_{-0.25}^{+0.21}$, $0.52_{-0.19}^{+0.19}$, for GW190706, GW190519, GW190620, GW170729, GW190517, respectively \citep{2021PhRvX..11b1053A}. We also note that these high values of $\chi_{\rm eff}$ are heavily under debate \cite[see e.g.,][references therein]{2022ApJ...937L..13C,2022arXiv220906978V}.

Substantial progress for understanding the origin of BBHs has been made in the field over the last 7 years since the discovery of the first GW event GW150914 \citep{2016PhRvL.116f1102A}. However, the formation process of BBH merger events remains an open scientific question. Leading models of BBH formation include isolated binary evolution via either common envelope \citep[CE, e.g.,][]{1991ApJ...380L..17P,1973NInfo..27...70T,2007ApJ...662..504B,2013A&ARv..21...59I,2014LRR....17....3P,2016Natur.534..512B,2018MNRAS.481.4009V,2018A&A...616A..28Q,2020A&A...635A..97B,2022ApJ...928..163H}, stable Roche-lobe overflow \citep[RLOF, e.g.,][]{2017MNRAS.471.4256V,2017MNRAS.468.5020I,2021A&A...647A.153B,2021A&A...651A.100O,2021ApJ...921L...2O,2021ApJ...922..110G,2021A&A...650A.107M,2022ApJ...926...83T,2022ApJ...930...26S,2022ApJ...931...17V,2022arXiv220903385V}, or chemical mixing \citep{2016A&A...588A..50M,2016MNRAS.458.2634M,2016MNRAS.460.3545D,2016A&A...585A.120S,2020MNRAS.499.5941D,2021MNRAS.505..663R}, as well as dynamical assembly in globular clusters and galactic nuclear clusters \citep[e.g.,][]{2015PhRvL.115e1101R,2016ApJ...816...65A,2020ApJ...888L...3S,2021MNRAS.505..339M,2022ApJ...927..231F}, or efficient migration assisted in active galactic nuclei (AGN) disks \citep{2019ApJ...878...85S,2020MNRAS.498.4088M,2020ApJ...898...25T,2022MNRAS.tmp.2697S}. Alternatively, two BHs can be the occurrence of hierarchical stellar-mass BH mergers \citep{2020ApJ...893...35D,2020ApJ...900..177K,2021ApJ...915L..35K,2021NatAs...5..749G}.

\citet{2021ApJ...910..152Z} recently investigated multiple formation pathways (isolated binary evolution channels and dynamical assembly channels) and found that neither channel can contribute more than $\simeq 70\%$ of the BBHs reported in GWTC-2. Moreover, it was pointed out in \cite{2022PhR...955....1M} (also see \cite{2020FrASS...7...38M,2022LRR....25....1M}) that the merger rates for BBHs can vary by orders of magnitude for different formation scenarios. So far, it is still a challenge to quantitatively predict the properties of merging BBHs due to uncertain physics involved in single and/or binary evolution \citep{2010CQGra..27q3001A,2015ApJ...806..263D,2015ApJ...814...58D,2018MNRAS.480.2011G,2020MNRAS.493L...6T,2022MNRAS.tmp.1776B,2022ApJ...925...69B,2022A&A...657A.116P}. 

Alternatively, merging BBHs could be formed through the  double-core evolution. This scenario involving the CE phase has been recently investigated, focusing on low-mass He-rich stars leading to form double NSs \citep{2006MNRAS.368.1742D,2015ApJ...806..135H,2018MNRAS.481.4009V}. More massive stars with mass-ratio close to one at low metallicities evolving from ZAMS (Zero Age Main Sequence) in close binaries can undergo several stable mass transfer phases during core hydrogen burning (Case A mass transfer phase) and thus form double He-rich stars as potential progenitors of BBHs \citep[see Figure 3 in][]{2016A&A...588A..50M}. On the other hand, two massive stars could first evolve to form a close binary system of a He-star and a main-sequence companion star after the first mass transfer, and subsequently the second mass transfer from MS/giant star to the He-star leads to form massive He-rich binary stars in a short orbit.

For now most BBH systems reported by the LVK Collaboration are still consistent with zero BH spins. Recently, by employing a variety of complementary methods to measure the distribution of spin magnitudes and orientations for BBH mergers, \cite{2022ApJ...937L..13C} found that the existence of a subpopulation of BHs with vanishing spins is not required by current data. The fact at the moment no event necessarily requires a high spin does of course not mean that there are not among those already detected any that may present a high spin. High BH spins may indicate that the inefficient AM transport mechanism within the BH progenitor is preferred \citep{2019IAUS..346..426Q,2019ApJ...870L..18Q,2022ApJ...924..129Q}. This finding can be reached given the assumption that BBHs are formed through the classical isolated binary evolution channel involving CE phase, before which the initially more massive star collapses to form the first-born BH. Accordingly, the progenitor of the first-born BH is in a wide orbit in which the tides from its companion are too weak to change the spin AM of both components. Therefore, the resultant BH spin, inherited from the AM content of its progenitor, is exclusively determined by the AM transport efficiency within the progenitor star during post main sequence expansion. In case of an efficient transport, any removal of the outer layers (at the time of CE phase) slows the whole star, even its core. In case of a less efficient coupling, the core spins faster than the envelope and removing the envelope will make appear a faster rotating core than in the case of the efficient AM transport. Alternatively, it is shown in \cite{2021ApJ...921L...2O} that fast-spinning BHs in merging BBHs can be formed by tidal spin-up through either a stable mass transfer phase leading to the mass ratio reversal, or the CE phase forming equal-mass BH components. For the case of stable mass transfer \cite[see their Figure 1 in][]{2021ApJ...921L...2O}, the initially more massive star evolves first to become a BH, and then its companion obtains enough mass via the first RLOF to become a massive He-rich star due to losing its hydrogen envelope onto the first-born BH in the second RLOF. The He-rich star subsequently evolves to become a fast-spinning BH by the tides \citep{2018A&A...616A..28Q}. As for the other case \cite[see their Figure 2 in][]{2021ApJ...921L...2O} the two stars initially with equal-mass instead form twin-mass He-rich stars following the RLOF mass transfer and subsequent CE phase, after which two fast-spinning BHs are formed via Wolf-Rayet tides. More recently, under the assumption of the Eddington-limited accretion onto BHs and efficient AM transport within massive stars, \cite{2022ApJ...933...86Z} investigated the isolated binary evolution regarding forming highly-spinning BHs and concluded that it is difficult to form systems with moderate or high spins in the primary BH component. However, the BH can be efficiently spun up by highly super-Eddington accretion \citep{2021A&A...647A.153B,2020ApJ...897..100V,2022RAA....22c5023Q,2022ApJ...930...26S}.

The Tayler-Spruit dynamo \citep{2002A&A...381..923S}, produced by differential rotation in the radiative layers, is considered as one of potential mechanisms responsible for the efficient transport of AM between the stellar core and its radiative envelope. In brief, the TS dynamo starts for a small radial magnetic field component (its precise initial value has no importance since it is rapidly enhanced by the dynamo mechanism). This component is wounded up per differential rotation and an azimuthal component field is formed. An azimuthal field is unstable by the Tayler instability, i.e., an nonaxisymmetric pinch type instability, which has consequence to amplify the azimuthal field and the radial one. The new radial component is wounded up and the instability starts again. This amplification mechanism lasts until the growth timescale of the magnetic field is equal to its damping timescale. Assuming that stationary situation is reached at every time step and that the length over which the instability can develop is small enough for allowing the excess energy in the differential rotation to overcome the stabilizing entropy gradient and large enough for the magnetic field to not decay too fast, it is possible to deduce the diffusion and viscosity coeffiecients. The revised TS dynamo \citep{2019MNRAS.485.3661F} is based on the fact that the damping timescale can be much longer than the one assumed in the original Tayler-Spruit dynamo. In that case larger magnetic fields can be reached and stronger coupling achieved \cite[see the discussion in][]{2022A&A...664L..16E}.

Stellar models with the original Tayler-Spruit dynamo (TS dynamo) can well reproduce the rotation rates for the Sun \citep{2005A&A...440L...9E}, white dwarfs and NSs \citep{2005ApJ...626..350H,2008A&A...481L..87S}. However the TS dynamo is currently challenged for explaining the slow rotation rates of cores in red giants \citep{2012A&A...544L...4E,2014ApJ...788...93C}. 
Recently, the revised TS dynamo \citep{2019MNRAS.485.3661F}, which was proposed to better match lower core rotation rates for sub-giant and red giant stars in better agreement with observed values, faces a challenge to reproduce the observational constraints on asteroseismic data of evolved stars \citep{2019A&A...631L...6E,2020A&A...634L..16D}. Applying the revised TS dynamo to massive He-rich stars in close binary systems predicts lower BH spins when compared with the original TS dynamo \citep{2022MNRAS.511.3951F}. More recently,  \cite{2022A&A...664L..16E} derived a new calibrated version of the original TS dynamo to better account for the evolution of the core rotation rates along the red giant branch stars when compared with the revised dynamo version. There was a theoretical debate on the existence of the dynamo \citep{2007A&A...474..145Z}. \cite{2022arXiv220908104J} recently performed three-dimentional magnetohydrodynamic simulations of the Tayler instability in rotating stellar interiors, and claimed to observe dynamo action via the amplification of poloidal magnetic field, indicating the TS instability could be important for magnetic field generation and AM transport in the radiative regions of evolving stars. The detailed comparisons between different versions of TS dynamo are beyond the scope of this work. Therefore, we are focused on the impact of the original TS dynamo within massive He-rich stars on the spin of resultant BH and its comparison when the TS dynamo is not included.

In this paper, we systematically investigate an alternative evolutionary scenario to form BBHs from double He-rich stars, i.e., double-core evolution first proposed by \cite{1995ApJ...440..270B} who studied the formation of double NSss. In Section 2, we introduce the main methods used in the stellar and binary evolution models. We present our detailed results in Section 3. The conclusions and discussion are summarized in Section 4.

\section{Methods}
We use release 15140 of \texttt{MESA} stellar evolution code \citep{2011ApJS..192....3P,2013ApJS..208....4P,2015ApJS..220...15P,2018ApJS..234...34P,2019ApJS..243...10P} to perform all of the binary evolution calculations in this work. We adopt three different kinds of metallicities, $\mathrm{Z}=\mathrm{Z}_{\odot}, 0.1\mathrm{Z}_{\odot}, 0.01\mathrm{Z}_{\odot}$, where the solar metallicity is $\mathrm{Z}_{\odot}=0.0142$ \citep{2009ARA&A..47..481A}. We create He-rich stars at zero-age helium main sequence following the same method as in \cite{2018A&A...616A..28Q,2020A&A...635A..97B,2022ApJ...928..163H,2022arXiv220205892F}, and then relax the created He-rich stars to reach the thermal equilibrium when the ratio of the He-burning luminosity to the total luminosity $\geq$ 99$\%$. We model convection using the standard mixing-length theory \citep{1958ZA.....46..108B} with a parameter $\alpha = 1.5$ and semiconvection according to \cite{1983A&A...126..207L} with an efficiency parameter $\alpha_{\rm sc} = 1.0$. We adopt Ledoux convection criterion to treat the boundaries of the convective zones and consider the step overshooting as an extension given by $\alpha_p = 0.1 H_p$, where $H_p$ is the pressure scale height at the Ledoux boundary limit. The network of \texttt{approx12.net} is adopted for nucleosynthesis.

We treat rotational mixing and AM transport as diffusive processes \cite{2000ApJ...544.1016H}, including the effects of Eddington-Sweet circulations, the Goldreich–Schubert–Fricke instability, as well as secular and dynamical shear mixing. We include diffusive element mixing from these processes with an efficiency parameter $f_c = 1/30$ \citep{1992A&A...253..173C,2000ApJ...544.1016H}. We use the standard efficient AM transport mechanism \citep[e.g.,][]{1999A&A...349..189S,2002A&A...381..923S}. Stellar winds of He-rich stars are modeled with the standard ``\texttt{Dutch}'' scheme, multiplied with a scaling factor of 2/3 to match the recently updated modeling of helium stars' winds \citep{2021MNRAS.505.4874H}. 

He-rich stars are modeled to reach the carbon exhaustion in the center. The baryonic remnant mass is calculated following the ``\texttt{delayed}'' supernova prescription as in \cite{2012ApJ...749...91F}. In order to calculate the mass and spin of the BH, we follow the framework in \cite{2019arXiv190404835B}, which has been recently implemented in recent work \citep{2020A&A...635A..97B,2022ApJ...928..163H}. We take into account the neutrino loss as in \cite{2020ApJ...899L...1Z}. We adopt 2.5 $M_{\odot}$ as the maximum NS mass. As a comparison (see appendix A), we also considered BHs formed through direct core collapse without receiving any mass loss or natal kicks \citep{1999ApJ...522..413F,2008ApJ...682..474B}. Very recently, it was reported on VFTS 243 that an X-ray quiet BH was born with a negligible kick in a massive binary within the Large Magellanic Cloud \citep{2022NatAs.tmp..162S}. 

Tidal interaction in close binary systems plays a critical role in the evolution of the orbit and the internal AM for the two stellar components. In this work, we use the dynamical tides model \citep{1975A&A....41..329Z,1981A&A....99..126H} to calculate the synchronization timescale ($T_{\rm sync}$), which is dependent on the tidal coefficient $E_2$. The two He-rich stars are assumed to be non-rotating at ZAHeMS. The main reason is that He-rich stars can be quickly spun up in close orbits. For both He-rich (also H-rich) stars, \cite{2018A&A...616A..28Q} recently updated an approximate expression of $E_2$, mainly depending the convective core radius and the star's radius for a wide range of initial masses and evolutionary stages at different metallicities.

In this study, we are focused on detailed investigations of a parameter space study with various initial conditions of close double He-rich stars. We cover the initial masses of He-rich stars from 5 - 65 $M_{\odot}$, the initial orbital periods in a range of 0.1 - 6 days. We evolve two He-rich stars with equal mass at different initial metallicities assuming two different AM transport mechanisms.

\section{Results}
\subsection{Hertzsprung-Russell diagrams of single He-rich stars}
\begin{figure*}[h]
     \centering
     \includegraphics[width=0.99\textwidth]{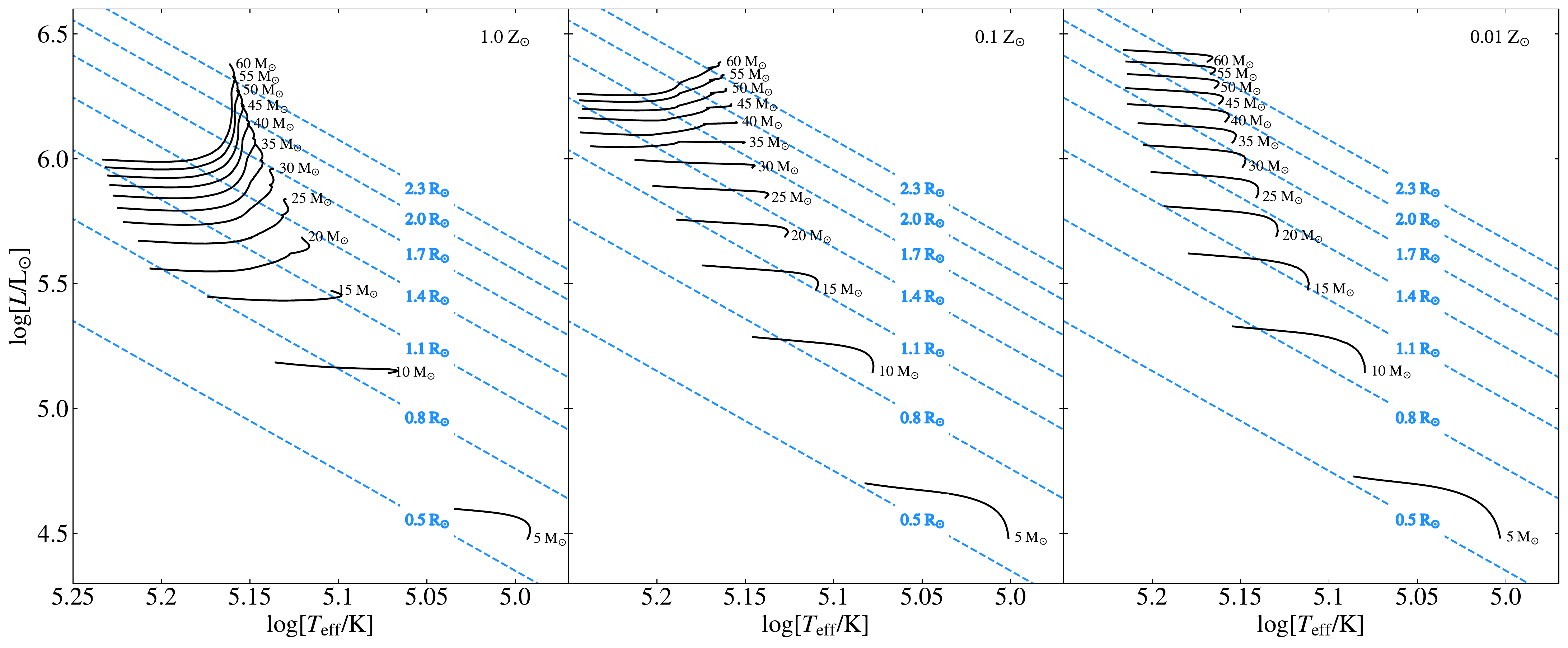}
     \caption{Hertzsprung-Russell diagrams of various single non-rotating He-rich stars with different initial metallicities (\textit{Left panel}: 1.0 $\mathrm{Z}_{\odot}$, \textit{middle panel}: 0.1 $\mathrm{Z}_{\odot}$. \textit{bottom panel}: 0.01 $\mathrm{Z}_{\odot}$.) evolving from Zero Age Helium Main Sequence (ZAHeMS) to the central helium exhaustion. The blue dashed lines refer to contours of constant radii.}
     \label{HRD}
\end{figure*}

Here we present the Hertzsprung-Russell (HR) diagram of single He-rich stars from the onset of the core helium burning (i.e., ZAHeMS) to the exhaustion of their central carbon. All of the He-rich stars are assumed to be non-rotating with different metallicities (1.0 $\mathrm{Z}_{\odot}$, 0.1 $\mathrm{Z}_{\odot}$ and 1.0 $\mathrm{Z}_{\odot}$), in the mass range of 5 - 60 $M_{\odot}$ at a step of 5 $M_{\odot}$. In Fig.~\ref{HRD}, the core helium burning phase begins on the right ends of the different curves labelled by the core He-mass, evolution then brings the stars to the left (the effective temperature increases). The evolution of the luminosity is different depending on the initial metallicity, rapidly decreasing at the beginning at 1.0 $\mathrm{Z}_{\odot}$ in the high mass range, and increasing at 0.01 $\mathrm{Z}_{\odot}$ in this same mass domain. This is an effect of the different mass loss rates at different metallicities. At high metallicities the strong mass loss rate decreases rapidly the luminoisity. At a low metallicity the mass is much less decreased and the main effect comes from the fact that the mean molecular weight increases increasing the luminosity, overcoming the effect due to the weak mass loss. These stars evolves towards to bluer regions of the HR diagram, which is similar to H-rich stars evolving chemically homogeneously on the main sequence. The main difference, however, is that for more massive He-rich stars their mass decreases and as a consequence the radius shrinks.

\subsection{Spin of BHs formed from double-core evolution}
\subsubsection{Impact of TS dynamo on BH spins}
Let us first show how different efficiencies of AM transport within He-rich stars change their rotation frequency at different evolutionary stages and thus the resulting spin parameters of BHs. As a case study, we evolve a binary system of two equal-mass He-rich stars, with initial mass $M_{\rm ZamsHe}$ = 39.80 $M_{\odot}$ at the initial orbital period $P_{\rm init.}$ = 0.63 days, until the end of their central carbon depletion.

First of all, we show in Fig.~\ref{omega} that the AM of the star and its core increases rapidly at the beginning due to the tidal interaction that spun up the star. Under the assumption that the wind mass lost is carrying the specific AM of the mass-losing star (Jeans mass loss), the He-rich star and its inner core will thus be slowed down. This situation, however, can be reversed for He-rich stars in a close binary system, in which tides are efficient to spin up the outer layers of the stars and their cores through strong coupling within the stars. We present the impact of the TS dynamo on the evolution of the internal rotation frequency of He-rich stars at different evolutionary stages, which are shown in the top two panels in Fig.~\ref{omega}. On the top left panel, we present models with the TS dynamo included (hereafter, TS on) given the solar metallicity, while we leave the discussion of a lower-metallicity model for the next section. First of all, with the TS dynamo the model in top left panel (see blue line) shows for the He-rich star in the middle of the core helium burning a flat distribution of a constant rotation frequency. This is because the star evolves with TS on like a solid body during core He burning phase. The whole star then gets spun up by the tidal interaction from its companion, as the star evolves off its core helium burning phase, from which on the star has rotation frequency of its outer layers slightly decreasing towards the surface due to the occurrence of increasing chemical gradient in the late evolutionary stage. The rotation frequency of the star continues increasing after the middle of the core helium burning, which is due to the tidal spun-up from its companion. In contrast, similar models without including TS dynamo (hereafter, TS off) on the top right panel in Fig.~\ref{omega}, show clear differences. The first difference is that the whole star due to less efficient coupling (e.g., TS off) between the outer layers and the stellar core is not a solid body from the early evolutionary stage (i.e., middle of core helium burning) to the late stages (see yellow sold line for the model at the central carbon ignition and red line for the central carbon depletion, respectively). Additionally, we also note that the star with TS off has a much larger rotational frequency throughout the whole evolutionary phase when compared with TS-on models. This is because inefficient coupling (TS off) between the outer layers and the stellar core allows the star to retain more AM and can instead be spun up if tides are strong.

In this section, we present the impact of TS dynamo on the evolution of the AM of the He-rich stars and their inner cores at different evolutionary stages. We show in Fig.~\ref{Jm10} that two binary evolutionary sequences of the same initial orbital period $P_{\rm init.}$ = 0.63 days and different initial masses $M_{\rm ZamsHe}$ = 10.00 $M_{\odot}$ (\textit{top row}) and $M_{\rm ZamsHe}$ = 39.80 $M_{\odot}$ (\textit{bottom row}), assuming TS on (solid lines) and off (dashed lines). Here we only show the solar-metallicity models, and leave the discussion of the rest in the following section. In the top left panel, we can see clear differences of the total AM of He-rich stars with TS on (black solid line) and off (black dashed line) starting before the middle of core helium (He) burning stage. For the model with TS on, during the core He burning phase the total AM of the star slowly decreases and then reaches the lower limit at the central He depletion. Additionally, the total AM keeps almost constant during the whole carbon (C) burning phase. Nevertheless, the model with the TS off shows that the star's total AM slightly decreases from the core He burning phase and then keeps constant until the central C depletion. The AM of the carbon-oxygen (CO) core of the He-rich star shows a similar trend, but with a shallow decay after igniting its the central carbon. The resulting BHs calculated using the prescription in \cite{2019arXiv190404835B} show the spins 0.08 (TS on) and 0.29 (TS off), respectively. For more massive He-rich binaries ($M_{\rm ZamsHe}$ = 39.80 $M_{\odot}$), the bottom left panel presents a similar finding, but with a much higher difference on the BH spin value, 0.07 (TS on) and 0.49 (TS off). First, He-rich stars in a very close binary are synchronised with their orbit due to strong tides, which allows more massive star to carry more AM given the same initial orbit when compared to binary systems of less components. On top of that, more massive He-rich stars are expected to have less lifetime before the core-collapse, resulting in more AM content within the progenitors and thus high resultant BH spins.

\subsubsection{Impact of the metallicity on BH spins}
As shown in the previous section, the TS dynamo has a significant impact on the evolution of the AM of the He-rich star and its core at later evolutionary stages, which further determines the spin values of the BH at birth. We here describe how the initial metallicity of He-rich stars can play a role in determining the spins of the resulting BHs. 

\begin{figure*}[h]
     \centering
     \includegraphics[width=0.99\textwidth]{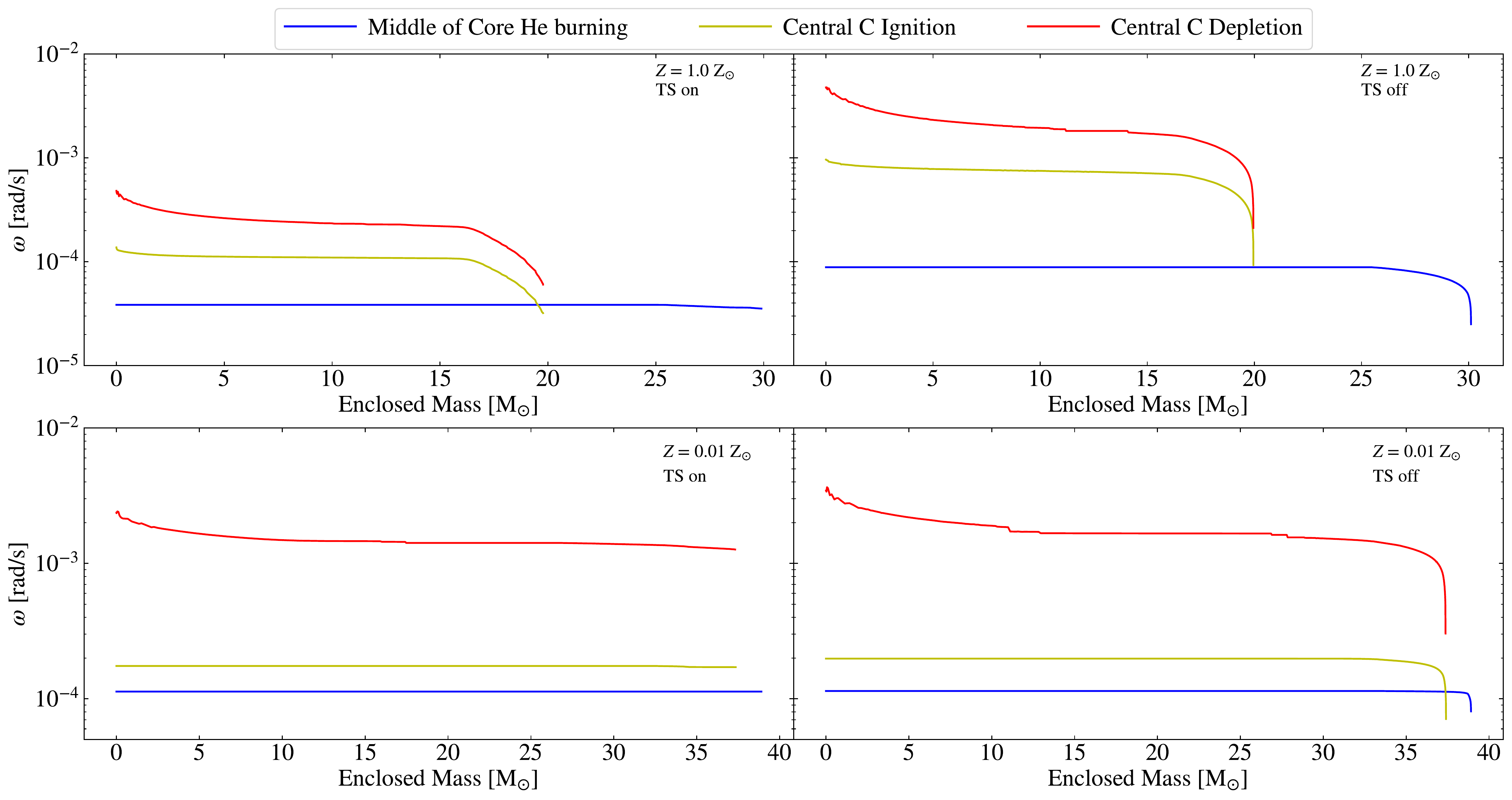}
     \caption{As a function of mass coordinate, we plot the angular velocity $\omega$ of He-rich stars at three evolutionary stages, i.e., middle of core helium burning (blue), Central carbon ignition (yellow), and central carbon depletion (red). The initial mass of He-rich star is $M_{\rm ZamsHe}$ = 39.80 $M_{\odot}$ and the initial orbital period $P_{\rm init.}$ = 0.63 days. Different efficiencies of AM transport mechanism and metallicities are assumed. \textit{Left column}: TS on, \textit{right column}: TS off. \textit{Top row}: 1.0 $\mathrm{Z}_{\odot}$, \textit{bottom row}: 0.01 $\mathrm{Z}_{\odot}$.}
     \label{omega}
\end{figure*}

It is well known that the stellar winds are strongly dependent on the metallicity of the mass-losing He-rich stars \citep{2001A&A...369..574V,2005A&A...442..587V,2006A&A...452..295E,2020MNRAS.491.4406S}. We can see that the He-rich star at its central carbon depletion has a much larger mass (around 37.5 $M_{\odot}$) at 0.01 $\mathrm{Z}_{\odot}$ when compared with a solar metallicity (around 20 $M_{\odot}$). We show in the two bottom panels of Fig.~\ref{omega} that, the He-rich star and its inner core have a similar and higher rotation rate at different evolutionary stages when compared with corresponding TS-on models at $\mathrm{Z}_{\odot}$. Additionally, the bottom right panel shows that the He-rich star evolves deviating from a solid body, slightly decreasing rotation rate from the stellar core to its outer layers. We also see that at solar metallicity the TS-off models retain more AM, but the difference is much less marked than at sub-solar metallicity (e.g., 0.01 $\mathrm{Z}_{\odot}$, see two bottom panels in Fig.~\ref{omega}). This difference is caused by the effect of the metallicity-depend wind mass loss which plays a critical in determining the final AM content of the progenitor star.

\begin{figure*}[h]
     \centering
     \includegraphics[width=0.99\textwidth]{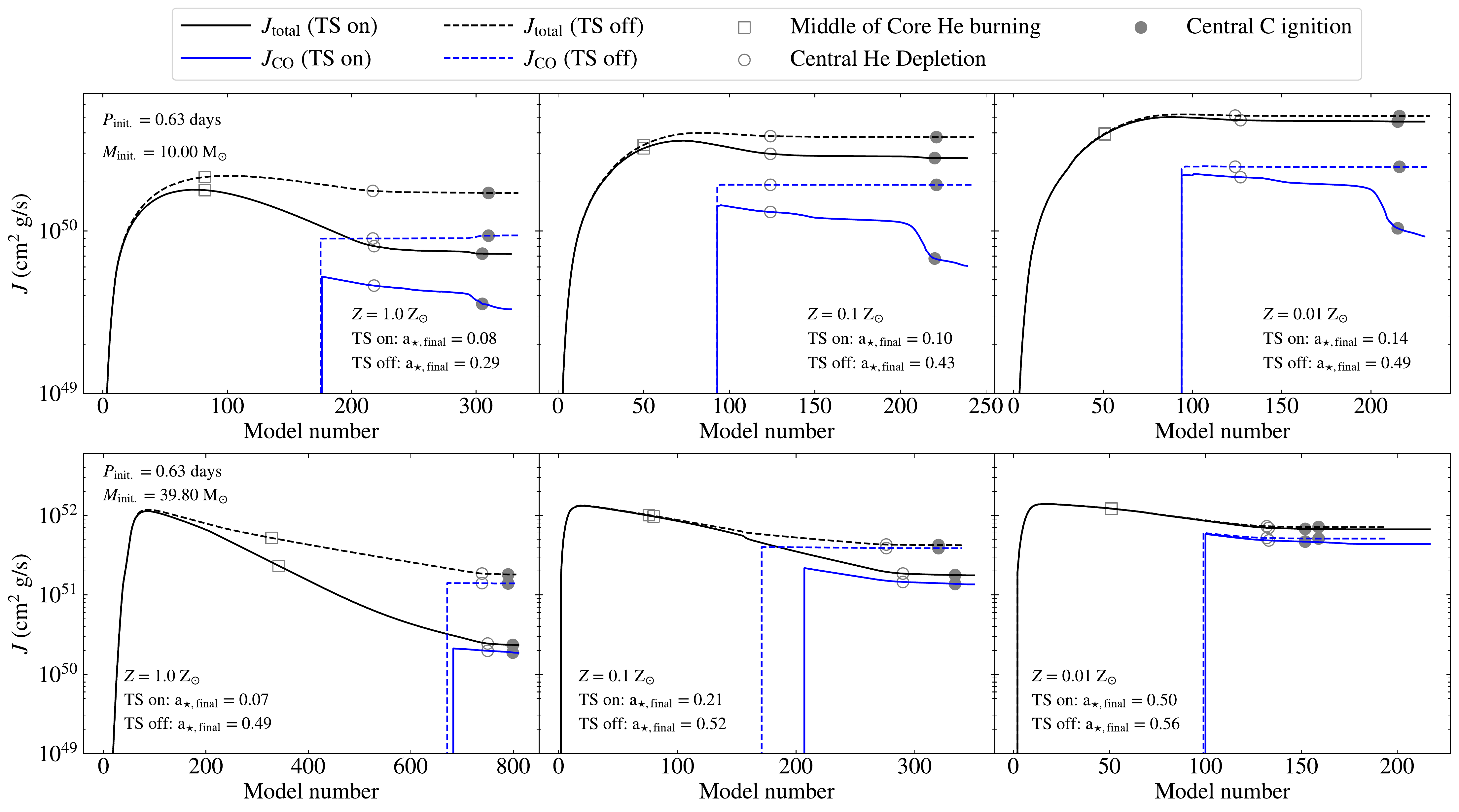}
     \caption{AM of the He-rich star and its Carbon-Oxygen (CO) core (black lines: $J_{\rm total}$, blue lines: $J_{\rm CO}$) as a function of model number for two binary sequences (\textit{Top row:} two equal-mass He-rich stars with initial helium star mass $M_{\rm ZamsHe}$ = 10.0 $M_{\odot}$, initial orbital period $P_{init.}$ = 0.63 days; \textit{bottom row:} $M_{\rm ZamsHe}$ = 39.8 $M_{\odot}$, $P_{init.}$ = 0.63 days). Similar to Fig.~\ref{omega}, we assume two different efficiencies of AM transport mechanism, i.e., solid lines: TS on, dashed lines: TS off. \textit{Left column}: 1.0 $\mathrm{Z}_{\odot}$, \textit{middle column}: 0.1 $\mathrm{Z}_{\odot}$, \textit{right column}: 0.01 $\mathrm{Z}_{\odot}$. Three evolutionary stages are marked in different symbols: square: middle of core helium burning, circle: central helium depletion, filled circle: central carbon ignition. The spin parameters of BHs formed from He-rich stars are presented.}
     \label{Jm10}
\end{figure*}

In Fig.~\ref{Jm10}, we can see that at 0.1 $\mathrm{Z}_{\odot}$, both He-rich stars and their cores have a higher AM at different evolutionary stages when compared with 1.0 $\mathrm{Z}_{\odot}$ (see the top left panel in Fig.~\ref{Jm10}). Interestingly, we note that the TS dynamo plays a small role in determining the evolution of the AM of He-rich star and its core at a low metallicity (see the top middle and right panel). This is expected as the wind carrying the specific AM of the mass-losing He-rich star is weaker at lower metallicities, which weakens the effect of AM transport within stars. Therefore, the spin values of resultant BHs formed at lower metallicities are accordingly higher, i.e., 0.1 $\mathrm{Z}_{\odot}$ and 0.14 at 0.01 $\mathrm{Z}_{\odot}$ for TS on (TS off: 0.43 $\mathrm{Z}_{\odot}$ and 0.49 at 0.01 $\mathrm{Z}_{\odot}$).

  \subsubsection{Parameter space analysis}
\begin{figure*}[h]
     \centering
     \includegraphics[width=0.99\textwidth]{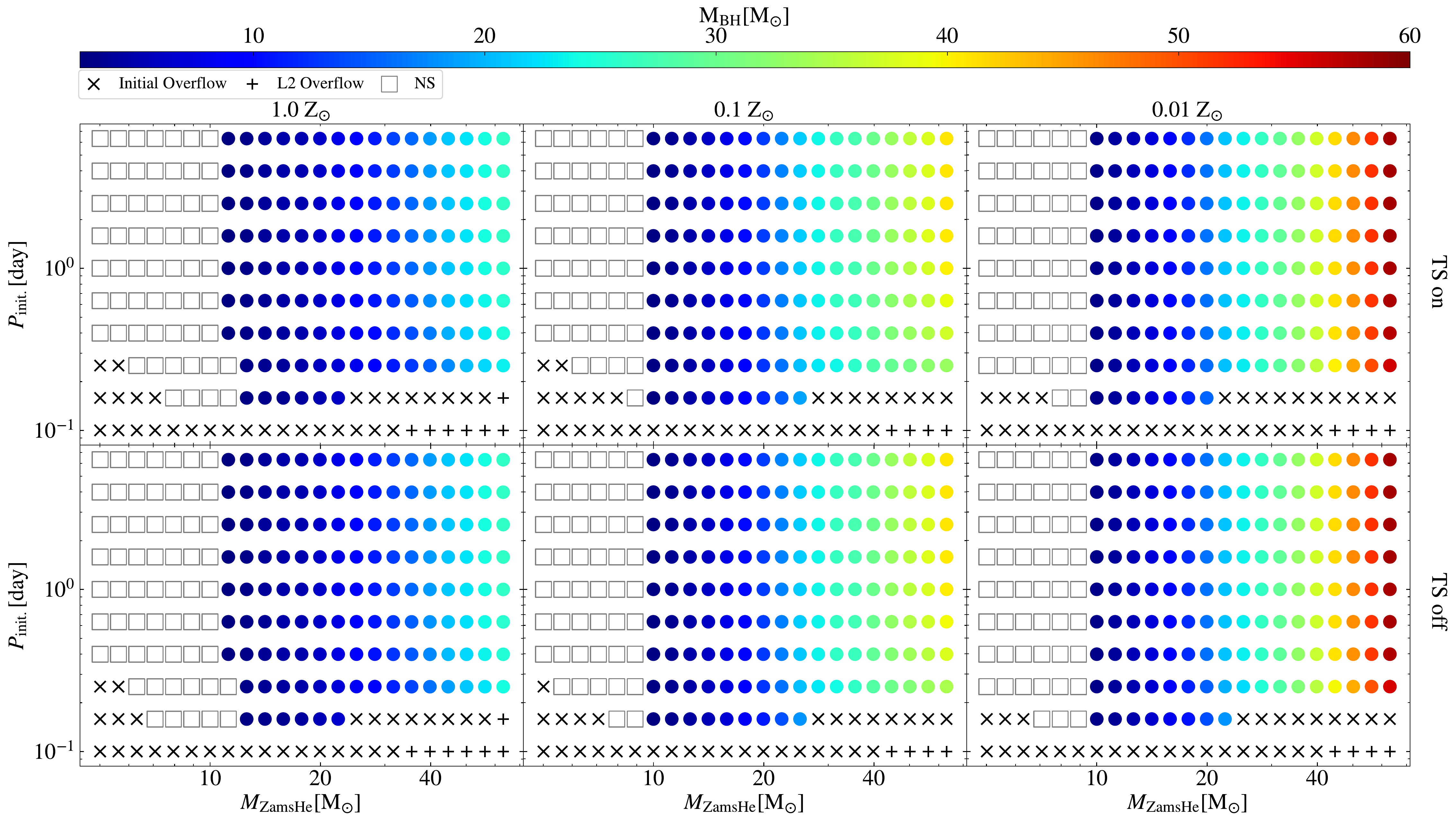}
     \caption{BH mass $\mathrm{M}_{\mathrm{BH}}$ as a function of the He-rich star's initial mass and orbital period, are marked with the color of the filled circles while gray squares represent that the compact objects formed through direct core-collapse of the He-rich star are NSs. \textit{Left column}: 1.0 $\mathrm{Z}_{\odot}$, \textit{middle column}: 0.1 $\mathrm{Z}_{\odot}$, \textit{right column}: 0.01 $\mathrm{Z}_{\odot}$. \textit{Top row}: TS on; \textit{bottom row}: TS off. The cross symbols represent the systems overflowing their Roche lobes at their initial models, while the plus symbols refer to the systems overflowing the second Legrangian point (L2) at the initial models.}
     \label{Final_m_Batta}
\end{figure*}

First of all, we show in the top left panel of Fig.~\ref{Final_m_Batta}, BH masses as a function of the He-rich initial mass and initial orbital period at solar metallicity. First, the binary systems either start to overflow their Roche lobes at the first model for $P_{\rm init.}$ $\sim$ 0.1 days (0.2 days for $M_{\rm ZamsHe}$ $\sim$ 30 $M_{\odot}$) or undergo the second lagrangian point (L2) overflowing for $M_{\rm ZamsHe}$ $\sim$ 38 $M_{\odot}$ and $P_{\rm init.}$ $\sim$ 0.1 days. Second, given the ``\texttt{delayed}'' supernova prescription \citep{2012ApJ...749...91F} the lower mass limit of the He-rich star that can collapse to form a BH (given solar metallicity) is around 12 $M_{\odot}$, below which a NS is formed instead (The study of NS formation is not considered in this work). Third, a He-rich star can form a BH with the maximum mass of around 26 $M_{\odot}$. At 0.1 $\mathrm{Z}_{\odot}$ (see top middle panel), we note that $\sim$ 40 $M_{\odot}$ BH can be formed. Notably, the initial orbital period starts to have an impact on the mass of the resulting BH when its immediate progenitor (He-rich star) has an initial mass $\gtrsim$ 40 $M_{\odot}$. This is because He-rich stars tend to lose more masses at a higher rotation rate in a closer binary system, which is due to the rotationally-enhanced mass loss \citep{1997ASPC..120...83L,2000A&A...361..159M}. It is clearly shown at 0.01 $\mathrm{Z}_{\odot}$ that more massive BHs ($>$ 55 $M_{\odot}$ see top right panel) can be formed. It is worth noting that the efficient AM transport within He-rich stars plays a negligible role in determining the BH mass.

\begin{figure*}[h]
     \centering
     \includegraphics[width=0.99\textwidth]{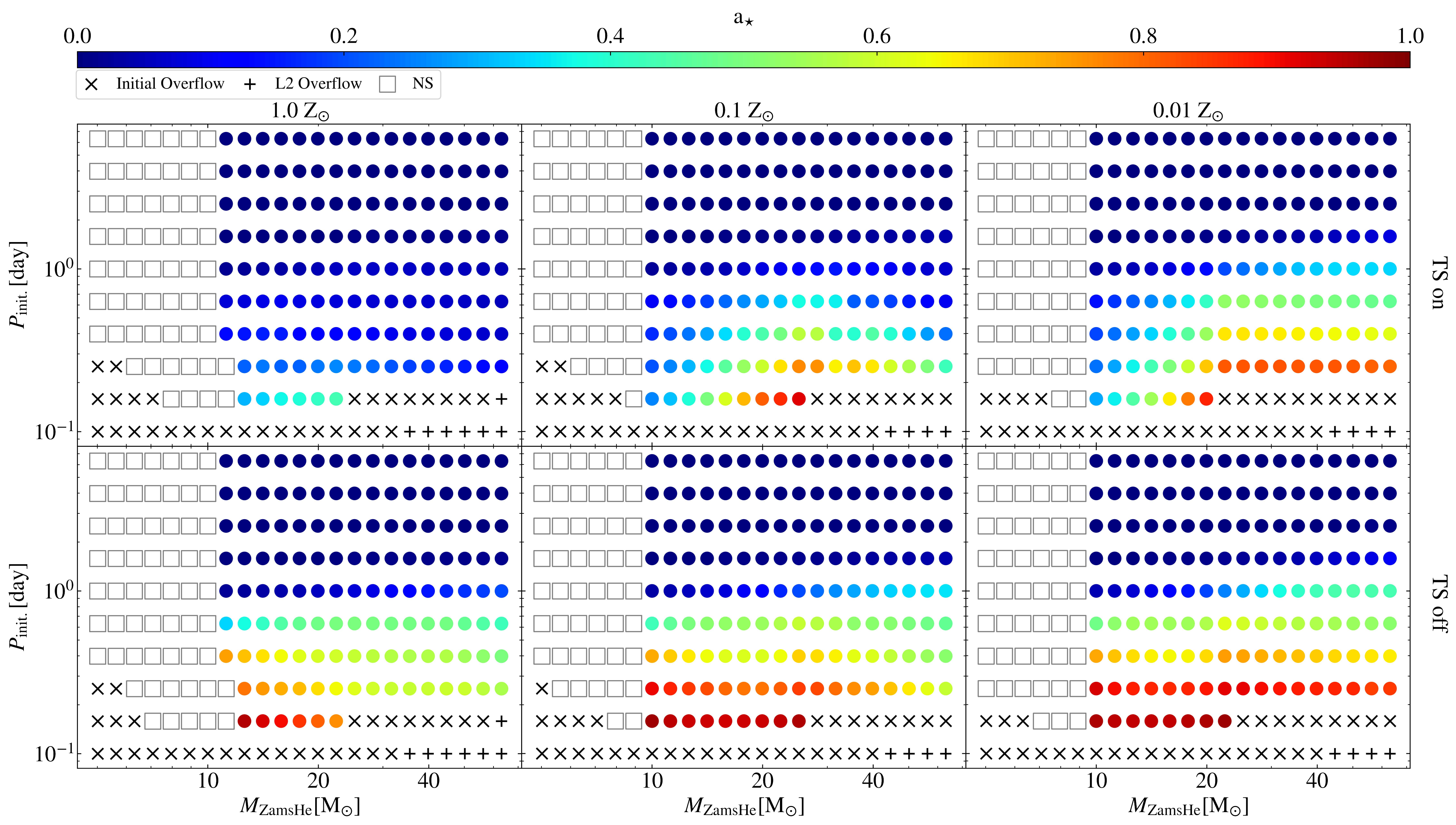}
     \caption{As in Fig.~\ref{Final_m_Batta}, but the color denotes the BH spin parameter $\mathrm{a}_{\star}$.}
     \label{Final_spin_Batta}
\end{figure*}

We then present in Fig.~\ref{Final_spin_Batta} the spin parameters $a_*$ of BHs formed from collapsing He-rich stars in close binaries with various conditions and assumed AM transport processes. Let us first show the spins of resultant BHs assuming efficient AM transport within He-rich stars. We note that the tides start to play a role when the initial orbital period $P_{\rm init.}$ is not longer than 1.0 day for all metallicities. As demonstrated in recent studies \citep{2018A&A...616A..28Q,2022MNRAS.511.3951F}, the interplay between the tides and wind mass loss of He-rich stars determines the AM of resultant BHs at birth and thus their spin magnitudes. At solar metallicity, the formed BHs are found to have low spin values (i.e., $a_*$ $\lesssim$ 0.4, see top left panel). This is because the wind mass loss of He-rich stars at a high metallicity is dominant over the tides. The spin magnitudes of BHs formed at lower metal poor environments are shown in the middle (0.1 $\mathrm{Z}_{\odot}$) and right panel (0.01 $\mathrm{Z}_{\odot}$). Therefore, at a given initial orbital period ($P_{\rm init.}$ $\lesssim$ 1.0 day), high BH spins can be reached for models at 0.1 and 0.01 $\mathrm{Z}_{\odot}$. At 0.01 $\mathrm{Z}_{\odot}$, we clearly see in the top right panel that, the spin magnitudes continue increasing with initial orbital period for all different initial masses of He-rich star. This is because the progenitor of the BH has very weak winds at 0.01 $\mathrm{Z}_{\odot}$, and thus loses negligible mass and AM. Furthermore, it is clear to see that the spins of BHs, originated from He-rich stars with initial mass $\lesssim$ 20 $M_{\odot}$, slightly increase with initial mass. This is because the wind of low-mass He-rich stars at very low metallicity (0.01 $\mathrm{Z}_{\odot}$) is significantly weak. Accordingly, for initially higher mass of He-rich stars, more infalling mass with its corresponding AM can be accreted to the newly-formed BHs \citep{2019arXiv190404835B}, resulting in higher final BH spins.

Assuming inefficient AM transport within He-rich stars, we show the spins of resultant BHs in the \textit{second row} of Fig.~\ref{Final_spin_Batta}. As shown clearly in the bottom left panel, high BH spins ($> 0.9$) can be reached at solar metallicity. Additionally, the spin covers the whole range, i.e., from minimum to maximum. For initial orbital periods $P_{\rm init.}$ $\lesssim$ 1.0 day, we note that the BH spin gradually decreases with increasing initial mass of He-rich stars, which is because massive He-rich stars are prone to be slowed down due to their strong winds at high metallicity. This is in contrast to the results of models at very low metallicity (see the top right panel), where the wind mass loss of He-rich stars is significantly weak at 0.01 $\mathrm{Z}_{\odot}$. Notably, we can see that He-rich stars tend to form higher-spinning BHs at lower metallicities which correspond to weaker wind mass loss (see bottom middle and right panel). 

\subsection{Merging timescales and comparisons with observed merging BBHs}
After two BHs form from the core-collapse of He-rich stars, gravitational wave (GW) emission shrinks the separation by removing the orbital AM, and eventually leads to the merger the compact objects. The timescale for two point masses to spiral in through GW emission from an initial eccentricity being zero (circular orbit) is given by \cite{1964PhRv..136.1224P}
    \begin{equation}\label{eq11}
    \centering
    T_{\rm merger} = \frac{5}{512}\frac{c^5}{G^3M^3}\frac{2 q^{-2}}{1+q^{-1}}a^4,
    \end{equation}
where $M$ is the BH mass, $q$ the mass ratio of the two BHs ($q = 1$ for our case) and $a$ is the orbital separation.

\begin{figure*}[h]
     \centering
     \includegraphics[width=0.99\textwidth]{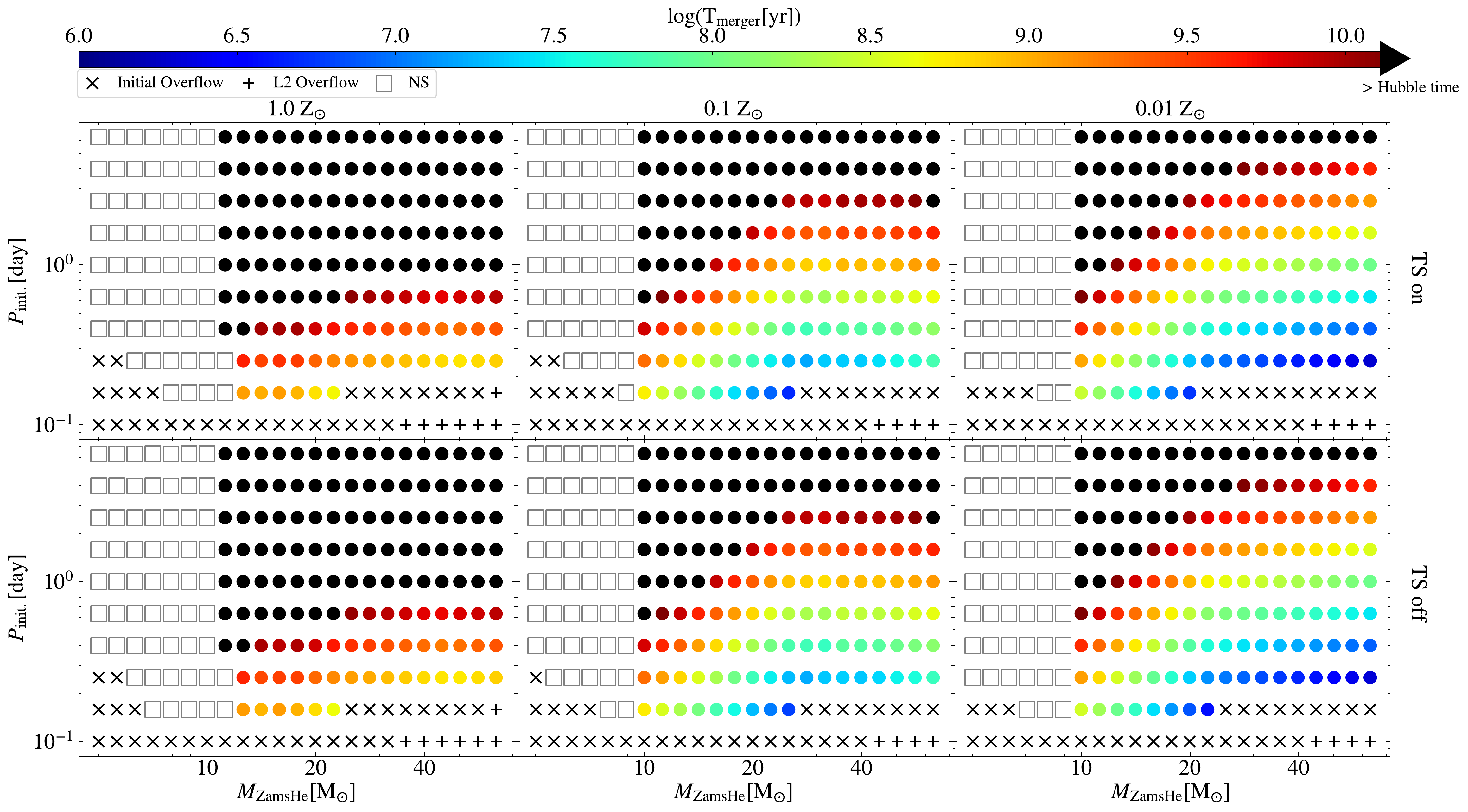}
     \caption{As in Fig.~\ref{Final_spin_Batta}, but the color represents merger time $\mathrm{T}_{\mathrm{merger}}$. The black dots represent the systems whose merger times are longer than a Hubble time}
     \label{mergertime}
\end{figure*}

We show the color bar in Fig.~\ref{mergertime} corresponding to $T_{\rm merger}$ of merging BBHs due to GW emission. First, comparing the two rows of different AM transport mechanisms shows negligible impact on the merging timescale. This is because significant differences are only expected for the AM content of the BH progenitors, rather than
the properties (two component masses and the final separation) of the binary system just after the birth of two BHs. Second, the parameter space of systems that are able to merge within a Hubble time is extended in lower metallicities. This is because BH progenitors at a higher metallicity tend to lose more mass, and the BBHs at birth thus have larger separations ($T_{\rm merger} \propto a^4$). Third, given a specific initial orbital period and metallcity, BBHs with initially higher mass have shorter merging timescales ($T_{\rm merger} \propto M^{-3}$).

\begin{figure*}[h]
     \centering
     \includegraphics[width=0.99\textwidth]{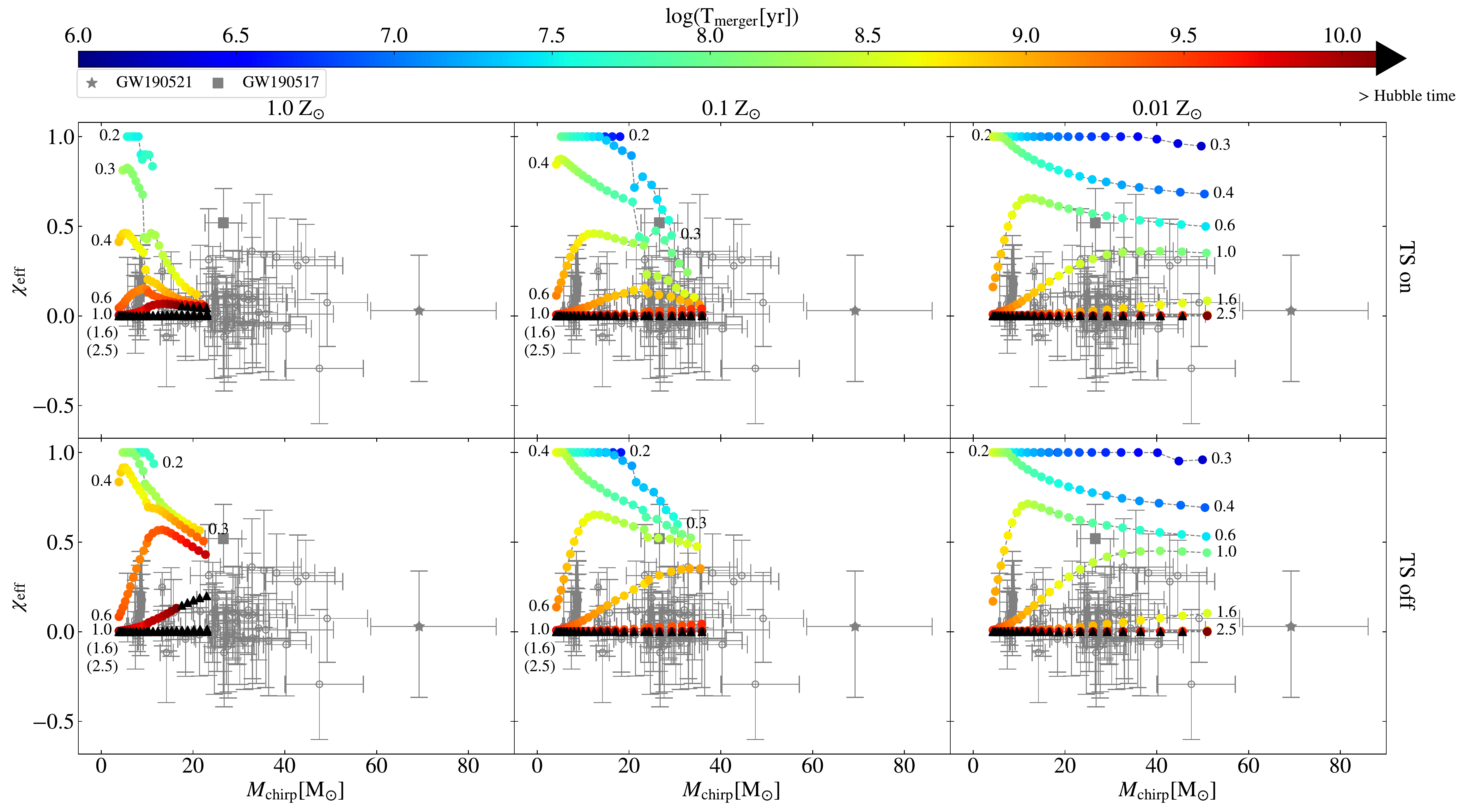}
     \caption{$T_{\rm merger}$ (colored dots) as a function of $\chi_{\rm eff}$ and $M_{\rm chirp}$. The \textit{first row} including three panels corresponds to different initial metallicities (\textit{left panel:} 1.0 $\mathrm{Z}_{\odot}$, \textit{middle panel:} 0.1 $\mathrm{Z}_{\odot}$, \textit{right panel:} 0.01 $\mathrm{Z}_{\odot}$) of He-rich stars. Similar to the \textit{first row}, but the \textit{second row} involves inefficient AM transport mechanism, i.e., TS off. The dashed lines refer to the specific initial orbital periods, marked with numbers in unit of ``days''. The observed high-confidence BBH events (69 BBHs with false alarm rate $<$ 1 per year) reported from the LVK are shown with the error bars at 90\% credibility in each panel. Black triangles represent BBHs whose $T_{\rm merger}$ is longer than a Hubble time. Two BBH events are highlighted, square: GW190517, star: GW190521.}
     \label{obs}
\end{figure*}

Figure ~\ref{obs} presents the merging timescales $T_{\rm merger}$ as a function of the effective inspiral spin $\chi_{\rm eff}$ and the chirp mass $M_{\rm chirp}$. 69 high-confidence BBH events (false alarm rate $<$ 1 per year) officially reported from the LVK are also shown in grey for comparison in each panel. We present systems formed from the same initial orbital period in dashed line. We assume that the formed BHs have spin components perfectly aligned to the direction of the orbital AM. First of all, we note that the initial metallicity plays an important role in forming systems with the observable properties ($\chi_{\rm eff}$ and $M_{\rm chirp}$). i.e., lower metallicities corresponding to formed systems with higher $\chi_{\rm eff}$ and larger $M_{\rm chirp}$. More specifically, the $M_{\rm chirp}$ can be reached around 26 $M_{\odot}$ at solar metallicity (40 and 58 $M_{\odot}$ at 0.1 and 0.01 $\mathrm{Z}_{\odot}$, respectively). Furthermore, the magnitude of $\chi_{\rm eff}$ can vary from 0.0 ($P_{\rm init.}$ = 1.0 day) and 1.0 ($P_{\rm init.}$ = 0.2 days). We note that so far no BBHs with $\chi_{\rm eff}$ = 1 has been reported from the LVK collaboration. Third, the AM transport mechanism in these observable properties of BBHs starts to play a more important role at higher metallicities (solar metallicity, see the two left panels in Fig.~\ref{obs}). Additionally, under the assumption of inefficient AM transport, double He-rich stars can form observable BBHs with $\chi_{\rm eff}$ > 0.80 (< 0.5 with TS on) with initially $P_{\rm init.}$ = 0.4 days and $\chi_{\rm eff}$ > 0.5 (< 0.25 with TS on) with initially $P_{\rm init.}$ = 0.6 days, respectively. It is also clearly shown in the left panels of Fig. ~\ref{obs} that more BBHs formed from double He-rich stars at solar metallicity will not be merged within a Hubble time (see black triangles) when compared with low-metallicity models. We note the trend that the observed BBHs with higher values of both $\chi_{\rm eff}$ and $M_{\rm chirp}$, can be better explained in our modeling at lower metallicites. In particular, the fraction of the BBHs, which have $\chi_{\rm eff}$ higher than that of GW190517 (it has the highest $\chi_{\rm eff}$ reported in the LVK), is 8.9\% at 1.0 $\mathrm{Z}_{\odot}$, 20.3\% at 0.1 $\mathrm{Z}_{\odot}$, and 26.9\% at 0.01 $\mathrm{Z}_{\odot}$ (For TS off: 18.1\% at 1.0 $\mathrm{Z}_{\odot}$, 23.2\% at 0.1 $\mathrm{Z}_{\odot}$, and 28.7\% at 0.01 $\mathrm{Z}_{\odot}$), respectively. GW190521 was reported from the LVK to have the highest $M_{\rm chirp}$ \citep{2020PhRvL.125j1102A}, which might be a straddling binary using a population informed prior \citep{2020ApJ...904L..26F}. This event is an outlier in our models, as the upper limit of the BH mass in this study is assumed not to be higher than $\sim$ 65 $M_{\odot}$ due to (pulsational) pair-instability supernovae (see discussion in the next section). Additionally, GW190517$_{-}$055101 has the largest $\chi_{\rm eff}$ reported in the second Gravitational-Wave Transient Catalog (GWTC-2) \citep{2021PhRvX..11b1053A}, which can be explained with our models at lower metallicities (see the second and third rows in Fig.~\ref{obs}), regardless of the assumed efficiencies of AM transport. Therefore, the BBH progenitor of this event might have gone through the double-core evolution at low metallicities (e.g., $Z <$ 0.1 $\mathrm{Z}_{\odot}$.)

\section{Conclusions and Discussion}
In this work, we first present the Hertzsprung-Russel diagrams of single non-rotating He-rich stars in a mass range of 5 - 60 $M_{\odot}$ at different metallicities, evolving from ZAHeMS to the central helium exhaustion. We then systematically study an alternative formation scenario of BBHs (i.e., the double-core evolution) by modeling double He-rich stars in various parameter spaces (metallicity and initial mass of He-rich stars, as well as the orbital period). Furthermore, we also investigate the impact of the different AM transport mechanism on the evolution of He-rich stars in different evolutionary stages, the properties of resulting BBHs at birth, as well as the merging timescale. 

We calculate the baryonic remnant mass following the ``\texttt{delayed}'' supernova prescription shown in \cite{2012ApJ...749...91F}, and taking into account the impact of accretion feedback onto the newly-formed BHs. The upper limit of the BH mass is around 26, 40 and 58 $M_{\odot}$ at 1.0 $\mathrm{Z}_{\odot}$, 0.1 $\mathrm{Z}_{\odot}$ and 0.01 $\mathrm{Z}_{\odot}$, respectively. We find that tides for double He-rich stars can only be important when the initial orbital periods are less than 1.0 day, which is similar to previous studies of a He-rich star accompanied by a BH/NS \citep{2018A&A...616A..28Q,2020A&A...635A..97B,2022MNRAS.511.3951F}. We note that the initial metallicity of He-rich stars should be high for the efficient AM transport to play a significant role in determining the spin magnitude of the newly-formed BHs, since their progenitors (massive He-rich He-stars) are more inclined to be slowed down by stronger winds mass-loss especially when rotating like a solid body. The $\chi_{\rm eff}$ for BBHs formed through the double-core evolution is not always high, but it can cover the whole range of BH spin, i.e., from minimum (0.0) to maximum (1.0), depending on the initial orbital period of the binary systems. The chirp mass $M_{\rm chirp}$ of BBH is strongly dependent on the initial metallicity of He-rich stars \citep[e.g.,][]{2010ApJ...715L.138B,2017NatCo...814906S,2019ApJ...882..121S}. More specifically, the chirp mass $M_{\rm chirp}$ of the BBH from double-core evolution at 1.0 $\mathrm{Z}_{\odot}$ can not be larger than 26 $M_{\odot}$, regardless of the efficiency of the AM transport within He-rich stars. 

After detailed investigations of the double-core evolution, we would expect that this channel could predict a certain fraction of BBH populations with high $\chi_{\rm eff}$ and $M_{\rm chirp}$. More events with the above features are expected to be captured by the LVK with its improving sensitivity in the upcoming fourth observing run. The quantitative merger rate from this channel is beyond the scope of the current work. Therefore we plan to investigate the quantitative contribution of this channel to the intrinsic BBH population with the population synthesis study and the impact of different physical processes on the outcomes in the near future.

 The formation of massive He-rich binary stars might not involve the CE phase, in which the criteria for its occurrence are still under development. Recent investigations suggest that the BBHs merger rate from the CE channel might be overestimated in rapid population synthesis studies \citep[e.g.,][]{2017MNRAS.465.2092P,2021A&A...650A.107M,2021A&A...645A..54K,2021ApJ...922..110G,2021A&A...651A.100O}. Their studies indicate that stable mass transfer channel could be a dominant channel for the formation of merging BBHs \citep[e.g.,][]{2022ApJ...930...26S,2022arXiv220613842B}. \cite{2022ApJ...931...17V} recently found that stable mass transfer channel preferentially form BBH systems with more massive component BH masses. Furthermore, by varying the metallicity-dependent cosmic star formation history, \cite{2022arXiv220903385V} found the variations affect the slope of the high mass end of the BBH mass distribution, but have a slight impact on the CE channel. In addition, massive He-rich binary stars could be formed through the CHE channel. In this channel, the two massive stars initially evolve in a close orbit and thus have strong chemical mixing due to strong tides.

Here we briefly summarise some main uncertainties in our binary modeling. First, stellar wind mass loss is one of key uncertain physical processes in the evolution of massive stars, which can have a significant impact on the mass and the spin of resultant BHs. Second, it is unclear whether supernova kicks (natal kicks) are imparted onto BHs during the core-collapse process. BHs formed from direct core-collapse of massive stars were considered to receive no natal kick \citep{2008ApJ...682..474B}. Nevertheless, we note that a recent work by \cite{2011ApJ...742...81F,2022ApJ...938...66T} which argued that, rather than dynamical formation, isolated binary evolution can still explain the observed BBHs if BHs have spin-axis tossed due to the supernova kicks during their formation process in the core collapse of massive stars. The stellar evolution theory predicts a mass ``gap'' in the BH birth function caused by the (pulsational) pair-instability supernovae \citep{1964ApJS....9..201F,1967ApJ...148..803R,1967PhRvL..18..379B,1968Ap&SS...2...96F,2003ApJ...591..288H}, which is still uncertain and thus plays a critical role in determining the upper limit of the BH mass below the ``gap” \citep[see][and references therein]{2021ApJ...912L..31W}. The constraints from current observations of BBHs reported from the LVK are still weak due to a statistically small sample. Therefore, we expect the sample of BBH events with higher $\chi_{\rm eff}$ and $M_{\rm chirp}$ will be significantly expanded in the upcoming fourth run, which will be used to make stronger constraints on the supernova kicks during the formation process of BHs from massive stars.

\begin{acknowledgements}
Y.Q. acknowledges the support from the Doctoral research start-up funding of Anhui Normal University and from Key Laboratory for Relativistic Astrophysics in Guangxi University. This work was supported by the National Natural Science Foundation of China (Grant Nos. 12003002, 12192220, 12192221, 11863003, 12173010) and the Natural Science Foundation of Universities in Anhui Province (Grant No. KJ2021A0106). G.M. has received funding from the European Research Council (ERC) under the European Union's Horizon 2020 research and innovation programme (grant agreement No 833925, project STAREX). All figures were made with the free Python module Matplotlib \citep{Hunter2007}. 
\end{acknowledgements}

\bibliographystyle{aa}
\bibliography{ref}

\begin{appendix}
\section{Direct core collapse with mass and angular momentum conserved}
\begin{figure*}[h]
     \centering
     \includegraphics[width=0.99\textwidth]{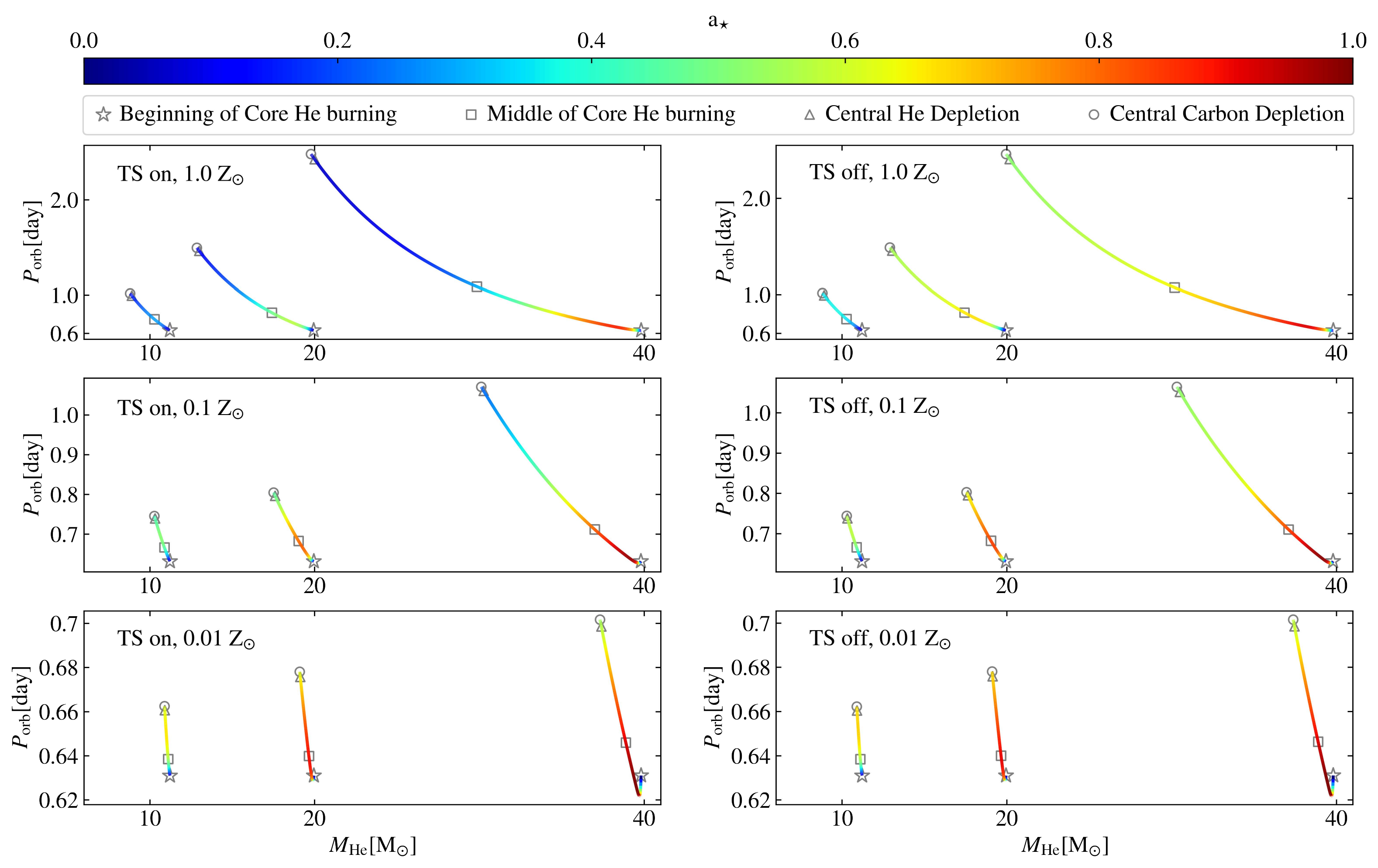}
     \caption{Evolution of the spin parameter $\mathrm{a}_{\star}$ as a function of the orbital period and mass of He-rich stars at different metallicities and AM transport mechanisms. \textit{Top left panel}: TS on and 1.0 $\mathrm{Z}_{\odot}$, \textit{middle left panel}: TS on and 0.1 $\mathrm{Z}_{\odot}$, \textit{bottom left panel}: TS on and 0.01 $\mathrm{Z}_{\odot}$; \textit{Top right panel}: TS off and 1.0 $\mathrm{Z}_{\odot}$, \textit{middle right panel}: TS off and 0.1 $\mathrm{Z}_{\odot}$, \textit{bottom right panel}: TS off and 0.01 $\mathrm{Z}_{\odot}$. The spin $\mathrm{a}_{\star}$ at different evolutionary stages are marked with symbols, star: beginning of core He burning, square: middle of core He burning, triangle: central He depletion, circle: central carbon depletion.}
     \label{Three_He_star}
\end{figure*}

In this section, we present the results of BBH formation through double-core evolution channel, assuming that the mass and AM are conserved during the formation process in the core collapse of He-rich stars. As shown earlier, tides can be only important for tidal interaction of double He-rich stars if the initial orbital periods are less 1 days.  We first show in Fig.~\ref{Three_He_star} the evolution of three cases ($M_{\rm ZamsHe}$ = 12, 20, and 40 $M_{\odot}$ for the same $P_{\rm init.}$ = 0.6 days) from the beginning of core helium burning to their carbon depletion in the center. We adopt efficient (TS on) and inefficient (TS off) AM transport within He-rich stars and three initial metallicities (1.0 $\mathrm{Z}_{\odot}$, 0.1 $\mathrm{Z}_{\odot}$ and 0.01 $\mathrm{Z}_{\odot}$). 

We show in the top left panel of Fig.~\ref{Three_He_star} the evolution of BH spin as a function of the He-rich star mass and its orbital period, under the assumption that He-rich stars at any time can directly collapse to form BHs without losing mass and corresponding AM. Let us begin with a case of an 40 $M_{\odot}$ double He-rich stars, it was efficiently spun up and thus formed a fast-spinning BH at the beginning of core helium burning (see the star symbol). The orbital separation slightly expands during the core helium burning, which however makes BH spin ($\sim$ 0.4 at the middle of core helium burning, see the square symbol) gradually decrease and end up with being close to zero at the central helium depletion. We note that there is a negligible discrepancy of BH spin calculated at between the central helium depletion (the triangle symbol) and the central carbon depletion (the circle symbol). The other two cases could form lower-mass binary BHs being slowly rotating in a closer binaries due to weaker winds mass loss. At lower metallcities, the same binaries will form faster-spinning BHs in shorter orbits (see middle left panel for 0.1 $\mathrm{Z}_{\odot}$ and bottom left panel for 0.01 $\mathrm{Z}_{\odot}$). When the inefficient AM transport (TS off) is adopted, we can clearly see the formed BBHs spinning faster when compared with the same metallicity.

\begin{figure*}[h]
     \centering
     \includegraphics[width=0.99\textwidth]{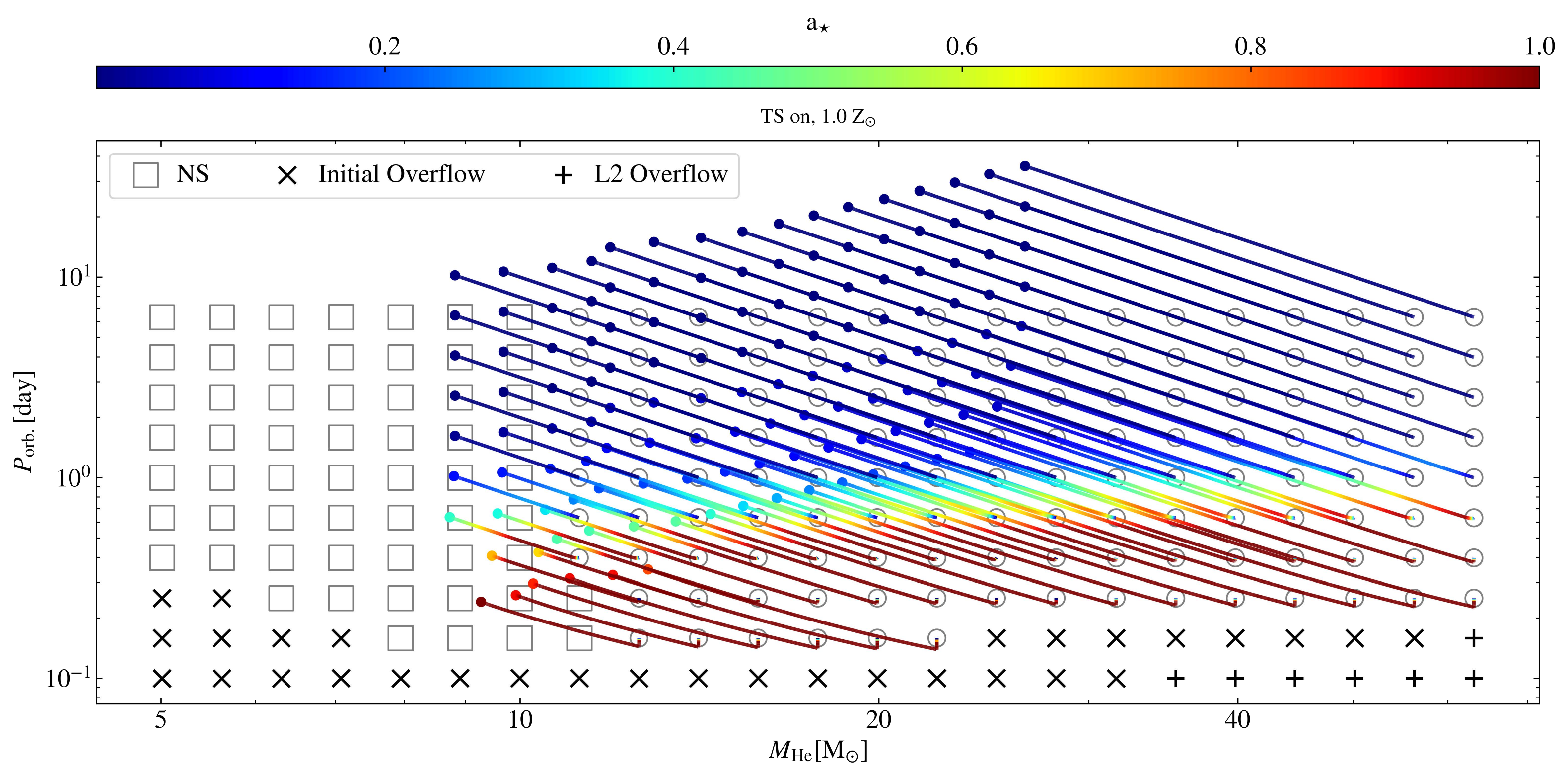}
     \caption{Evolution of the spin parameter $\mathrm{a}_{\star}$ (\textbf{the color bar}) as a function of the orbital period and mass of He-rich stars with TS included at solar metallicity. \textbf{The colored lines linking the two symbols show the evolution of the binary. The color along the line gives BH spins $\mathrm{a}_{\star}$ along the evolution (from the ZAHeMS to the central carbon depletion), assuming that their progenitors (He-rich stars) directly collapse to form BHs with mass and AM conserved.}}
     \label{fig4}
\end{figure*}

\begin{figure*}[h]
     \centering
     \includegraphics[width=0.99\textwidth]{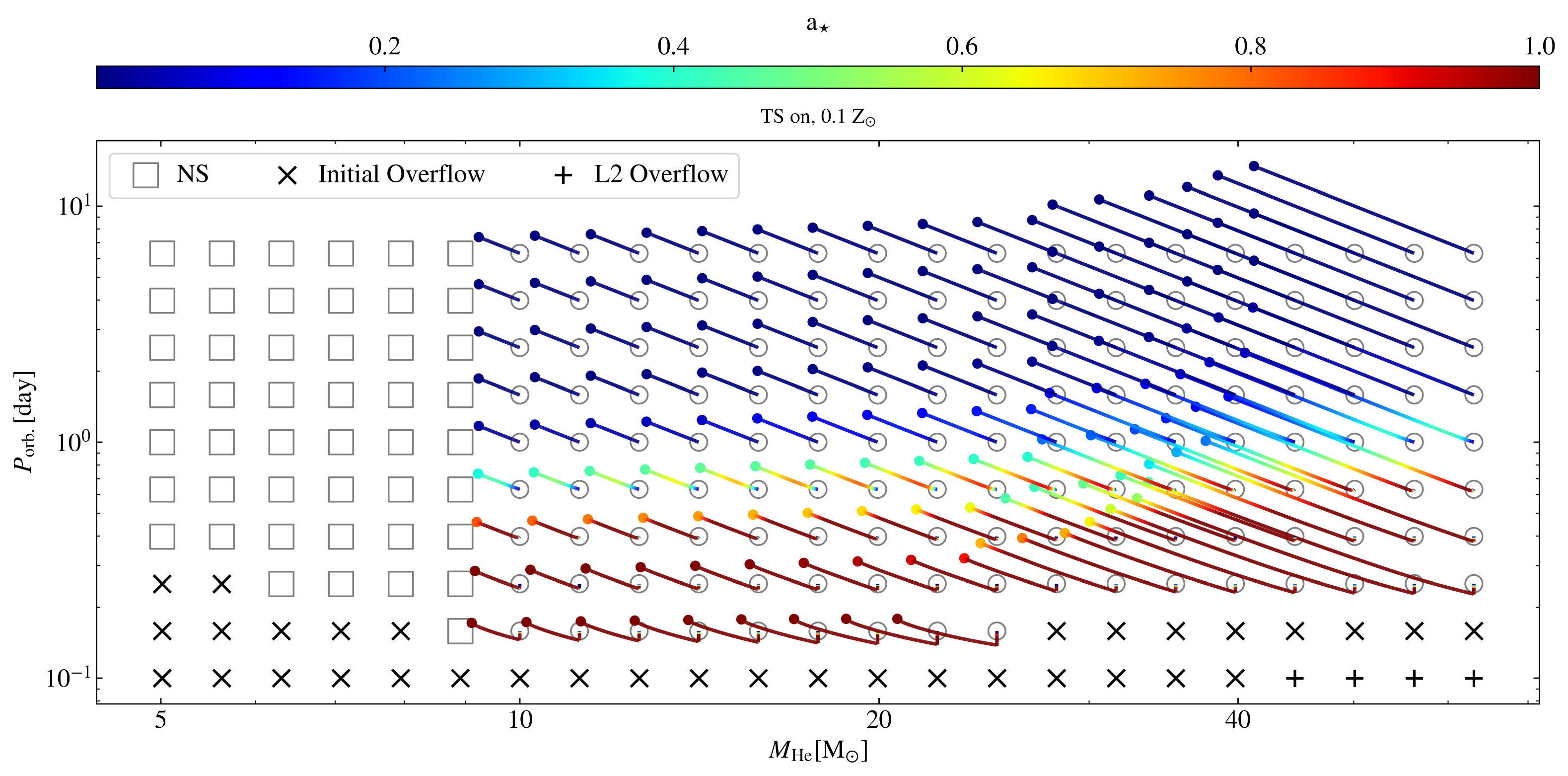}
     \caption{As in Fig .~\ref{fig4}, but for the metallicity $\mathrm{Z}=0.1\ \mathrm{Z}_{\odot}$.}
     \label{fig5}
\end{figure*}

With the same parameter space, we also compute for each binary system the evolution of the BH spin under different metallicities and efficiencies of AM transport. We first present results assuming efficient  AM transport. As shown in Fig.~\ref{fig4}, all He-rich stars with an initial mass of less than 12 $M_{\odot}$, at 1.0 $\mathrm{Z}_{\odot}$, form NSs. He-rich stars with initial orbital period of longer than 1.0 day end up with being non-spinning BHs. We find that BHs can have moderate spin magnitudes with $P_{\rm init.}$ in a range of 0.3 - 1.0 days, below which fast-spinning BHs are formed. Similar to Fig.~\ref{fig4}, we can see for lower metallicities (see Fig.~\ref{fig4} and Fig.~\ref{fig5}) that the formed BHs with spins decreasing as the orbit slowly expands. We then show the results with different metallicities of the inefficient AM transport in Fig.~\ref{fig7}, Fig.~\ref{fig8} and Fig.~\ref{fig9}. The mass and spin of the newly-formed BHs calculated using direct core-collapse with mass and AM conserved are slightly larger when compared with those by taking into account the accretion feedback during core-collapse modeling \citep[see details in][]{2019arXiv190404835B}.

\begin{figure*}[h]
     \centering
     \includegraphics[width=0.99\textwidth]{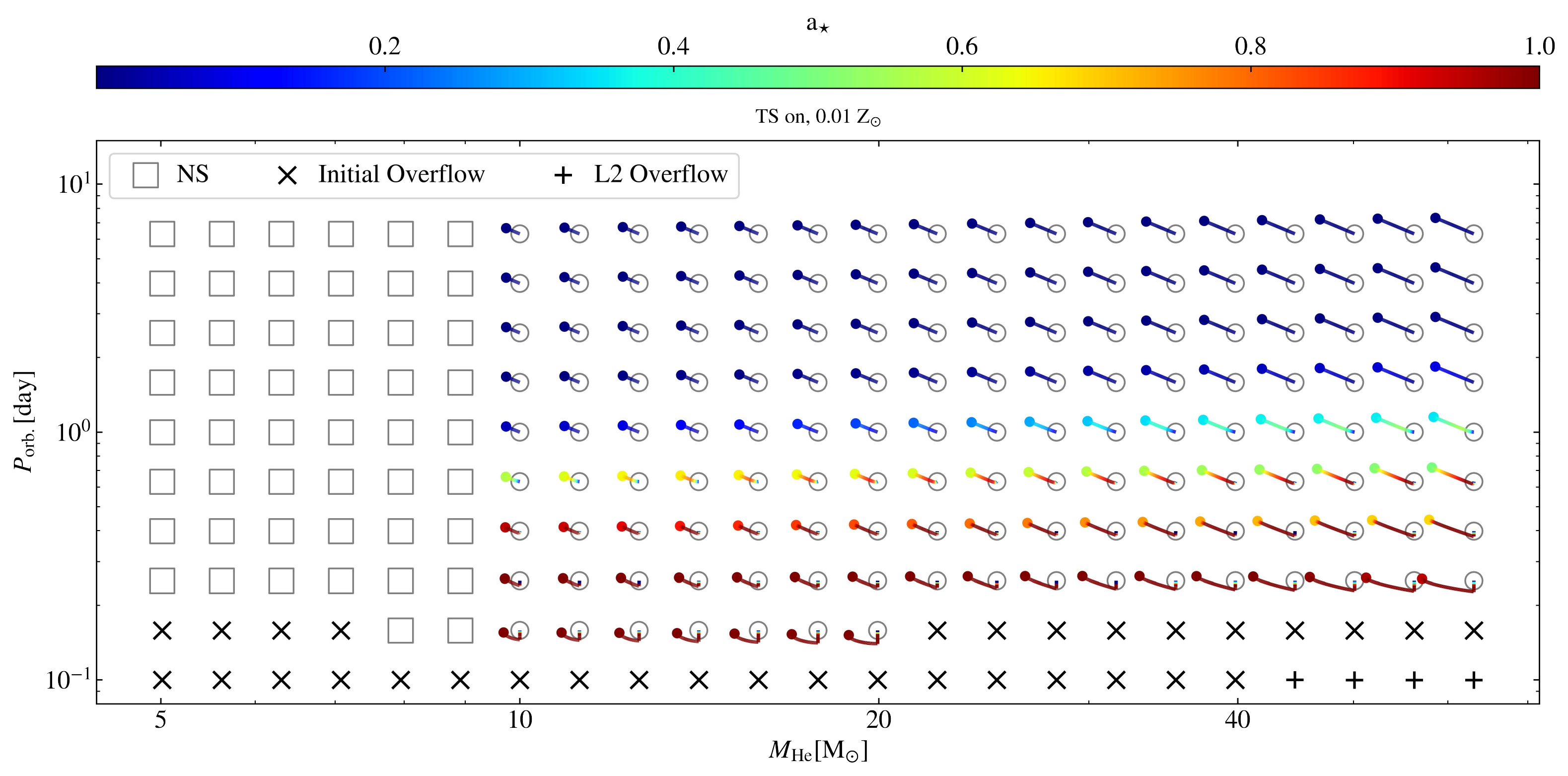}
     \caption{As in Fig .~\ref{fig4}, but for the metallicity $\mathrm{Z}=0.01\ \mathrm{Z}_{\odot}$.}
     \label{fig6}
\end{figure*} 

\begin{figure*}[h]
     \centering
     \includegraphics[width=0.99\textwidth]{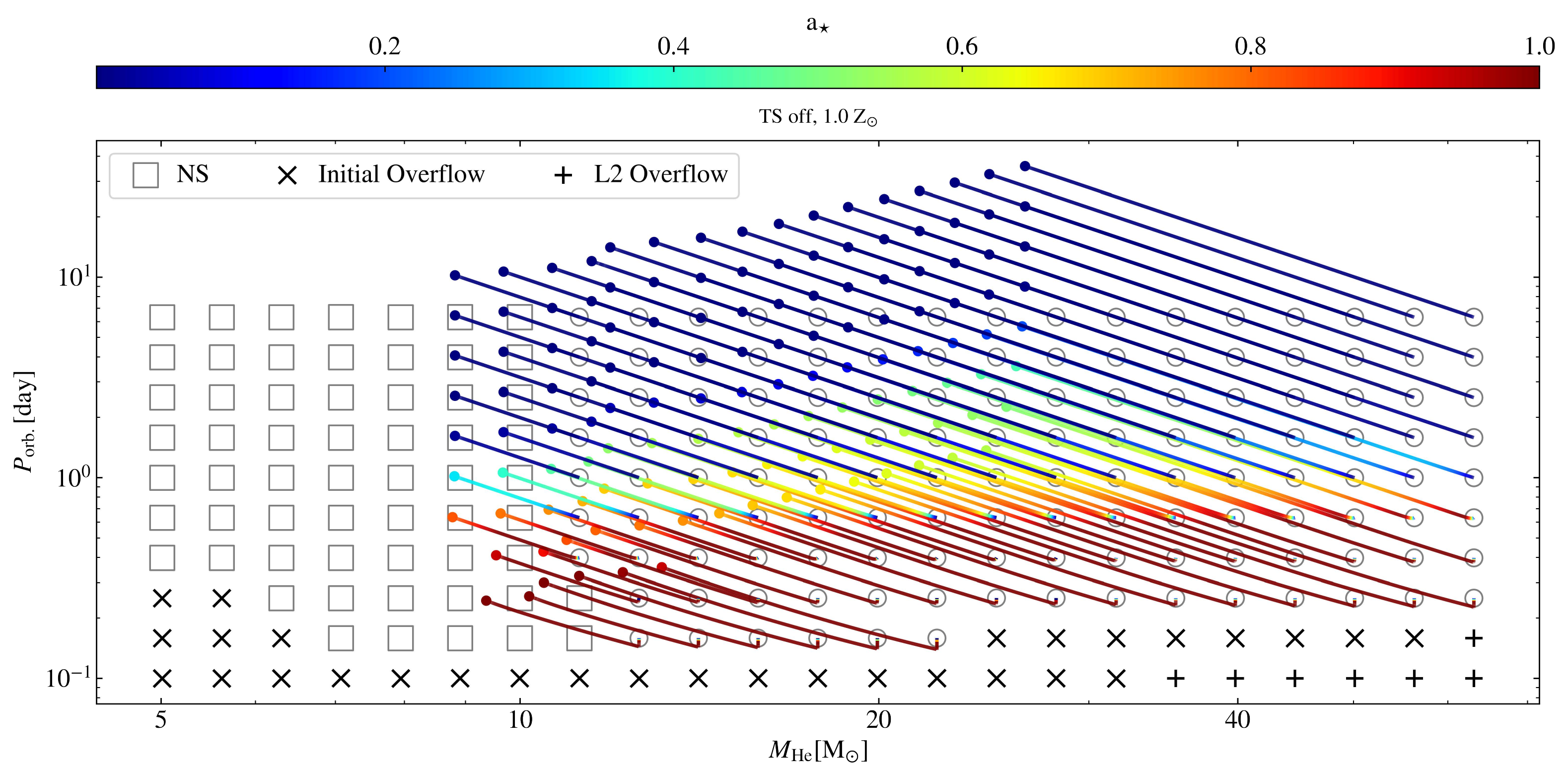}
     \caption{As in Fig .~\ref{fig4}, but without TS included.}
     \label{fig7}
\end{figure*} 

\begin{figure*}[h]
     \centering
     \includegraphics[width=0.99\textwidth]{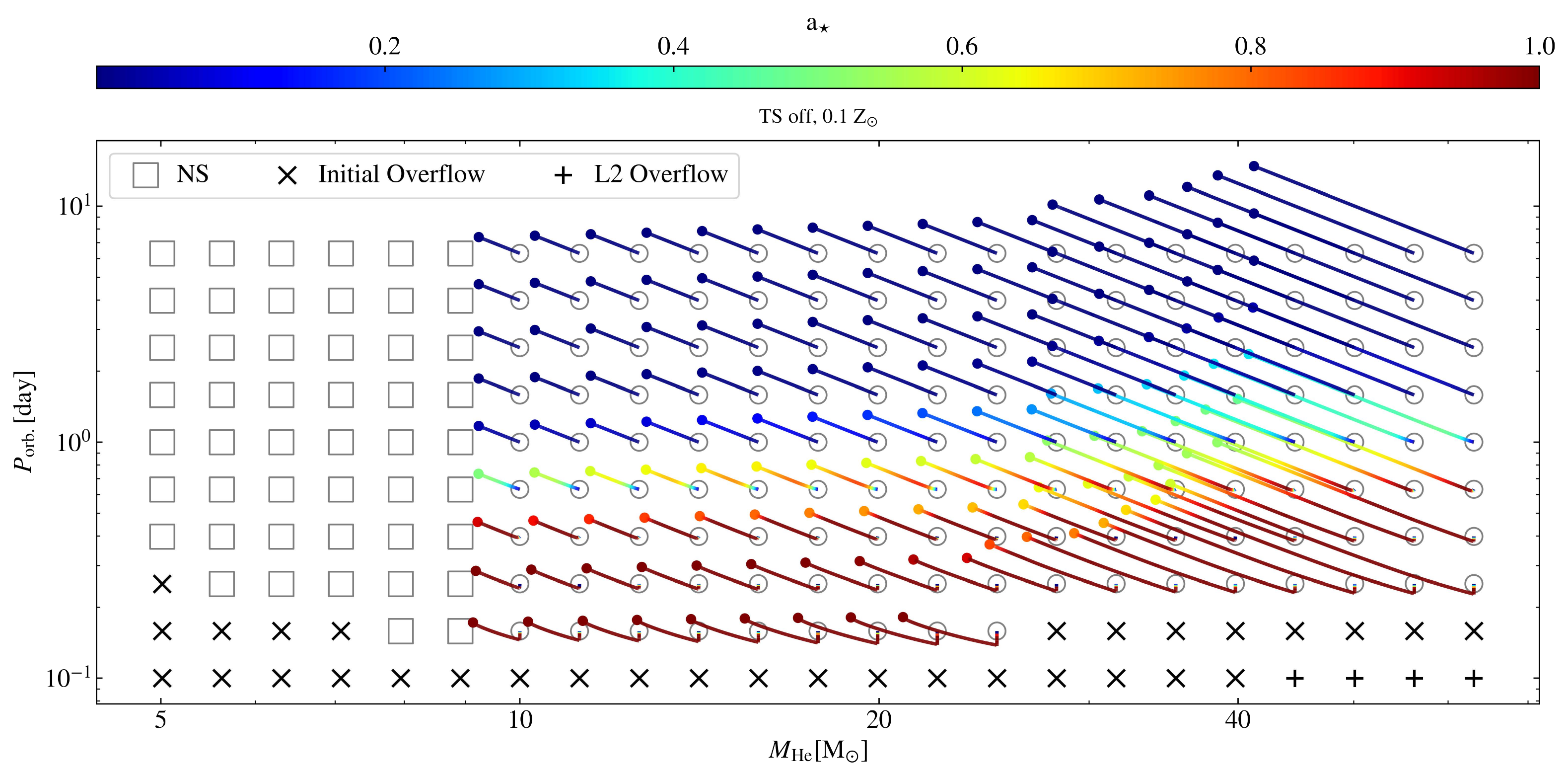}
     \caption{As in Fig .~\ref{fig7}, but for the metallicity $\mathrm{Z}=0.1\ \mathrm{Z}_{\odot}$.}
     \label{fig8}
\end{figure*} 

\begin{figure*}[h]
     \centering
     \includegraphics[width=0.99\textwidth]{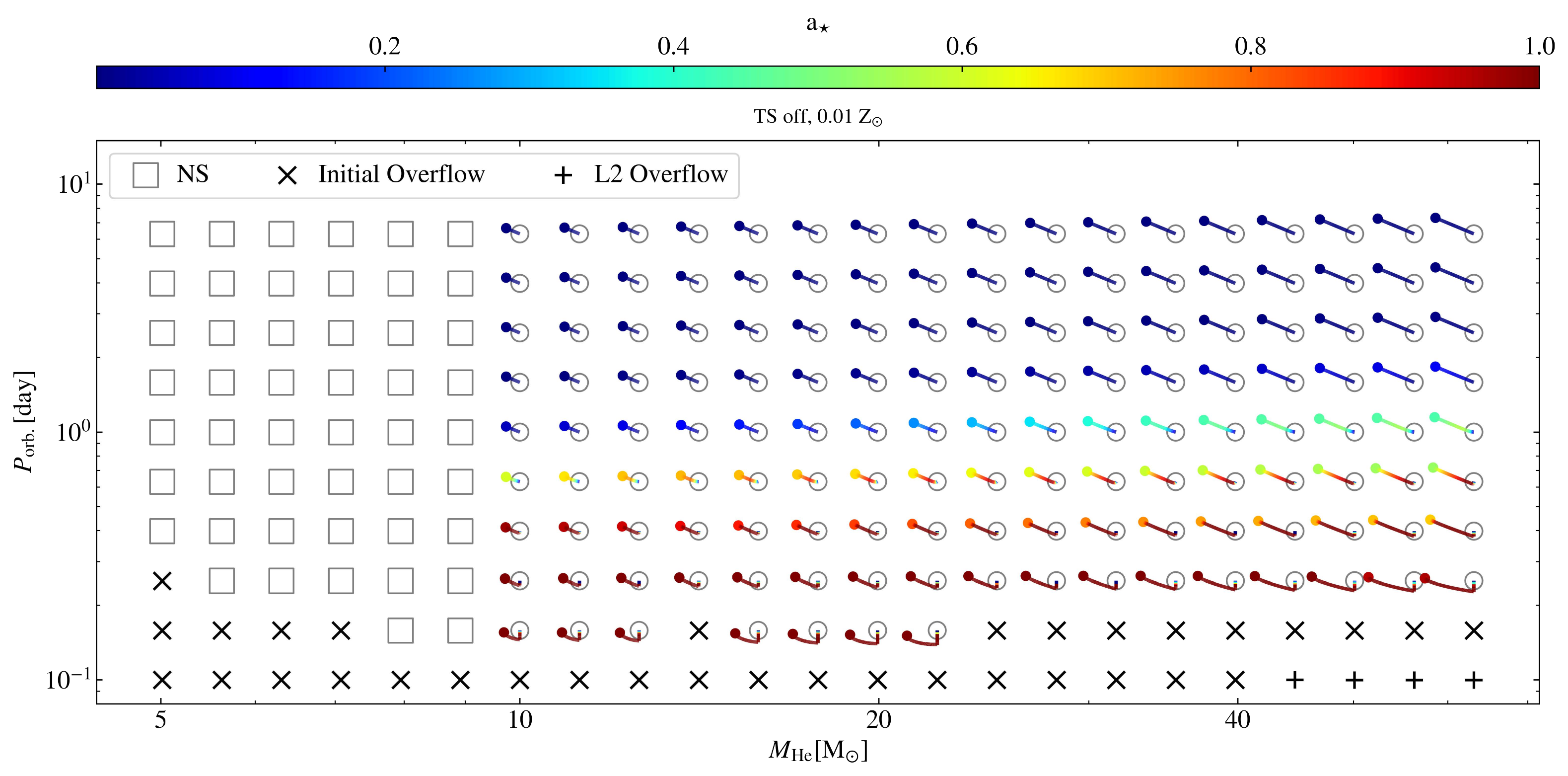}
     \caption{As in Fig .~\ref{fig7}, but for the metallicity $\mathrm{Z}=0.01\ \mathrm{Z}_{\odot}$.}
     \label{fig9}
\end{figure*} 
   
\end{appendix}

\end{document}